\documentclass[twocolumn,showpacs,preprintnumbers,amsmath,amssymb,floatfix]{revtex4}

\usepackage[normalem]{ulem}
\usepackage{verbatim}
\usepackage{graphicx}
\usepackage{dcolumn}
\usepackage{bm}
\usepackage{color}
\usepackage{url}
\usepackage{amsmath}
\usepackage{morefloats}
\usepackage{times}
\usepackage[varg]{txfonts}
\usepackage{multirow}
\usepackage{lineno}
\usepackage{hyperref}
\usepackage{notoccite}

\usepackage{color}

\newcommand{\MA}[1]{{\color{cyan}{#1}}}

\graphicspath{{./figures/}}
\newcommand{\be}{\begin{equation}}
\newcommand{\ee}{\end{equation}}
\newcommand{\ba}{\begin{eqnarray}}
\newcommand{\ea}{\end{eqnarray}}
\newcommand{\nn}{\nonumber}

\newcommand{\LL}{\mathcal{L}}
\def\ltsima{$\; \buildrel < \over \sim \;$}
\def\simlt{\lower.5ex\hbox{\ltsima}}
\def\gtsima{$\; \buildrel > \over \sim \;$}
\def\simgt{\lower.5ex\hbox{\gtsima}}
\begin{document}

\title[blah]{Parameterized tests of the strong-field dynamics of general relativity
using gravitational wave signals from coalescing binary black holes: Fast likelihood
calculations and sensitivity of the method}

\author{Jeroen Meidam$^{1}$, Ka Wa Tsang$^{1}$, Janna Goldstein$^{2}$, Michalis Agathos$^{3}$, 
Archisman Ghosh$^{1}$, Carl-Johan Haster$^{4}$, Vivien Raymond$^{5}$, Anuradha Samajdar$^{1}$, Patricia
Schmidt$^{6}$, Rory Smith$^{7}$, 
Kent Blackburn$^{8}$, 
Walter Del Pozzo$^{9}$, Scott E.~Field$^{10}$,
Tjonnie Li$^{11}$, Michael P\"{u}rrer$^{5}$, Chris Van Den Broeck$^{1,12}$, John Veitch$^{13}$, 
Salvatore Vitale$^{14}$}

\affiliation{$^1$Nikhef -- National Institute for Subatomic Physics, 105 Science Park, 
1098 XG Amsterdam, The Netherlands \\
$^2$School of Physics and Astronomy, University of Birmingham, Birmingham, B15 2TT, United Kingdom \\
$^3$DAMTP, Centre for Mathematical Sciences, University of Cambridge, Wilberforce Road, 
Cambridge CB3 0WA, United Kingdom \\
$^4$Canadian Institute for Theoretical Astrophysics, University of Toronto, Toronto, Ontario M5S 3H8, 
Canada \\
$^5$Albert-Einstein-Institut, Max-Planck-Institut f\"{u}r Gravitationsphysik, D-14476 Golm, Germany \\
$^6$Department of Astrophysics / IMAPP, Radboud University Nijmegen, P.O.~Box 9010, 6500 GL Nijmegen, 
The Netherlands \\
$^7$OzGrav, School of Physics \& Astronomy, Monash University, Clayton 3800, Victoria, Australia \\
$^8$LIGO, California Institute of Technology, Pasadena, CA 91125, USA \\
$^9$Dipartimento di Fisica ``Enrico Fermi", Universit\`{a} di Pisa, 
Pisa I-56127 and INFN sezione di Pisa, Italy \\
$^{10}$Mathematics Department, University of Massachusetts Dartmouth, Dartmouth, MA 02747, USA \\
$^{11}$Department of Physics, The Chinese University of Hong Kong, Shatin, NT, Hong Kong \\
$^{12}$Van Swinderen Institute for Particle Physics and Gravity, University of Groningen, \\
Nijenborgh 4, 9747 AG Groningen, The Netherlands \\
$^{13}$Institute for Gravitational Research, University of Glasgow, Glasgow G12 8QQ, United Kingdom \\
$^{14}$LIGO, Massachusetts Institute of Technology, Cambridge, MA 02139, USA}

\date{\today}

\begin{abstract}
Thanks to the recent discoveries of gravitational wave signals from binary black hole mergers 
by Advanced Laser Interferometer Gravitational Wave Observatory and Advanced Virgo, 
the genuinely strong-field dynamics of spacetime can now be probed, allowing for 
stringent tests of general relativity (GR). One set of tests consists of allowing for 
parametrized deformations away from GR in the template waveform models and then constraining 
the size of the deviations, as was done for the detected signals in previous work. 
In this paper, we construct reduced-order quadratures so as to speed up likelihood 
calculations for parameter estimation on future events. Next, we explicitly demonstrate 
the robustness of the parametrized tests by showing that they will correctly 
indicate consistency with GR if the theory is valid. We also check to what extent 
deviations from GR can be constrained as information from an increasing number of 
detections is combined. Finally, we evaluate the sensitivity of the method to 
possible violations of GR.
\end{abstract}


\date{\today}

\maketitle

\section{Introduction}

Since 2015, the twin Advanced LIGO 
observatories~\cite{TheLIGOScientific:2014jea} have routinely been detecting
gravitational wave signals from coalescing binary black holes 
\cite{Abbott:2017xlt,Abbott:2016nmj,TheLIGOScientific:2016pea,Abbott:2017vtc,Abbott:2017gyy},  
recently also in conjunction with Advanced Virgo 
\cite{TheVirgo:2014hva,Abbott:2017oio}. Later in the decade 
the worldwide gravitational wave detector 
network will be extended with the Japanese KAGRA \cite{Aso:2013eba}, to be followed by LIGO-India
\cite{Indigo}. In the 
course of the next several years, tens to hundreds more binary black hole detections
are expected to be made \cite{TheLIGOScientific:2016pea}. 

Coalescences of stellar mass binary black holes (BBHs) are ideal laboratories for testing the 
genuinely strong-field dynamics of general relativity (GR) \cite{Berti:2015itd,TheLIGOScientific:2016src}: 
They are likely to be pure 
spacetime events, involving stronger curvatures and shorter dynamical timescales
than in any other experiment or observation, by many orders of magnitude \cite{Yunes:2016jcc}. 
The process starts with two black holes that are orbiting each other, gradually 
losing orbital energy and orbital angular momentum through the emission of
gravitational waves (GWs). By the time the GW frequency is high enough to be in 
the sensitive band of Earth-based detectors, the binary will likely have shed 
almost all of its original eccentricity \cite{Peters:1963ux} and will be undergoing 
\emph{quasi-circular inspiral}.
Eventually the inspiral becomes non-adiabatic, after which the components of 
the binary undergo a \emph{plunge} followed by \emph{merger}, leading to the 
formation of a single, highly excited black hole. The latter undergoes 
\emph{ringdown} as it asymptotes to a quiescent, 
Kerr black hole. The GW signal that is emitted can, at large distances, be described as a 
small metric perturbation propagating at the speed of light on a Minkowski background; however,
the \emph{shape} of the wave encodes detailed information about the strong-field, 
dynamical inspiral-merger-ringdown (IMR) process it originated from. 

A number of techniques have been developed to understand 
inspiral-merger-ringdown in GR, including 
large-scale numerical relativity (NR) 
simulations resulting from direct integration of the Einstein equations
\cite{Pretorius:2005gq,Campanelli:2005dd,Baker:2006ha}, and
the construction of (semi-)analytic waveform models.
The effective one-body (EOB) formalism
\cite{BuonannoDamour:1999,BuonannoDamour:2000,Damour:2008qf,Damour:2009kr,Barausse:2009xi}
has been extended to combine the post-Newtonian (PN) description of inspiral \cite{Blanchet:2002av} with 
NR results for the merger, 
as well as black hole perturbation theory for the ringdown 
\cite{Vishveshwara:1970zz,Press:1971wr,Chandrasekhar:1975zz}, leading to high-quality 
IMR waveforms in the time domain \cite{Taracchini:2013rva}. In the frequency domain, 
phenomenological IMR models \cite{Ajith:2007kx,Ajith:2009bn,Santamaria:2010yb}
were developed based on a frequency domain PN expansion together with hybridized 
EOB/NR waveforms \cite{Husa:2015iqa,Khan:2015jqa,Hannam:2013oca}. 

A variety of possible deviations from GR have been considered in the context of 
binary coalescence, including scalar-tensor theories, a varying Newton constant,
modified GW dispersion relations, \emph{e.g.}~arising from ``massive graviton" models,
violations of the no-hair hypothesis, violations of Cosmic Censorship, and parity
violating theories (see \emph{e.g.}~\cite{Yunes:2016jcc} and references therein). 
Even within the GR paradigm, one can think of alternative compact objects
to black holes (\emph{e.g.}~boson stars, dark matter stars, or gravastars), 
which may exhibit tidal effects during inspiral (see \cite{Giudice:2016zpa,Cardoso:2017cfl}), 
and will also have a different 
ringdown signal from a black hole. 
For some alternative theories it has been worked 
out how the post-Newtonian inspiral would be modified to leading order \cite{YunesPretorius:2009}, and for 
certain exotic 
 objects the ringdown spectrum has been computed \cite{Cardoso:2016oxy}. 
However, what seems to be lacking in all 
cases are the kind of high-accuracy IMR waveforms that are available for BBH coalescence 
in GR. Thus, given observational GW data for a detected compact binary coalescence 
event, at present it is not possible to compare GR with alternative theories, or 
BBH coalescences with those of alternative compact objects, while making use of the 
full information in the IMR signal. Moreover, GR might be violated in an altogether 
different way that is yet to be envisaged.

Given these restrictions, at the present time it is expedient to devise tests of 
the theory of general relativity itself, which to the largest extent possible are generic, 
and as accurate as we can make them. 
Following the recent binary black hole merger detections, 
a battery of such tests were deployed 
\cite{TheLIGOScientific:2016src,TheLIGOScientific:2016pea,Abbott:2017vtc}: looking for 
coherent excess signal power in the 
data after subtraction of the best-fitting GR waveform \cite{LittenbergCornish2009,Cornish:2014kda}, 
checking for consistency with 
GR between the pre-merger and post-merger signals in terms of masses and spins
\cite{Ghosh:2016qgn,Ghosh:2017gfp}, evaluating
consistency of the post-merger signal with the presence of a least-damped ringdown mode 
\cite{TheLIGOScientific:2016src}, 
constraining anomalous GW propagation with a view on 
bounding the mass of the graviton as well as violations of local Lorentz invariance 
\cite{Will:1998,Mirshekari:2011yq} (the latter also using the binary neutron 
star detection \cite{TheLIGOScientific:2017qsa,Kostelecky:2016kfm,Monitor:2017mdv}), 
looking for evidence of non-standard polarization
states \cite{Will:2005va}, and measuring a 
series of judiciously chosen coefficients associated 
with parameterized deformations of IMR waveforms away from GR 
\cite{Blanchet:1994ex,Blanchet:1994ez,Mishra:2010tp,YunesPretorius:2009,Cornish:2011ys,Li:2011cg,Li:2011vx,Agathos:2013upa,TheLIGOScientific:2016src,TheLIGOScientific:2016pea,Abbott:2017vtc}. 
This paper deals primarily with the latter tests.

As mentioned above, a number of IMR waveform models have been developed. For parameterized 
tests of GR we use the phenomenological models, which have a closed expression in the 
frequency domain and hence can be generated fast on a computer (which is important for data 
analysis purposes when exploring high-dimensional parameter spaces), capture the essential 
physics of the problem (including \emph{e.g.}~spin-induced precession), and allow
for some amount of analytic insight into the meaning of the induced deformations. In particular, 
we use the model which in the LIGO Algorithm Library is designated as IMRPhenomPv2 
\cite{Husa:2015iqa,Khan:2015jqa,Hannam:2013oca}. 
The IMRPhenomPv2 waveform phase is characterized by a number of parameters $\{p_i\}$: 
(i) in the adiabatic inspiral regime, PN coefficients $\{\varphi_0, \ldots, \varphi_7\}$ and 
$\{\varphi_{5l}, \varphi_{6l}\}$; (ii) in the intermediate regime between adiabatic inspiral and merger, 
phenomenological coefficients $\{\beta_0, \ldots, \beta_3\}$, and (iii) in the merger-ringdown 
regime, $\{\alpha_0, \ldots, \alpha_5\}$. In the most relevant of these coefficients, parameterized 
deformations are introduced by allowing for relative deviations: 
$p_i \rightarrow (1 + \delta\hat{p}_i)\,p_i$. The $\delta\hat{p}_i$ will be referred to as
our \emph{testing parameters}.

We then perform a series of tests, in each of which some testing parameter  
$\delta\hat{p}_j$ is allowed to vary freely along with all other parameters entering the
phase (component masses and spins, which enter through the GR expressions for the $p_i$ 
themselves), but $\delta\hat{p}_k = 0$ for $k \neq j$. In principle multiple $\delta\hat{p}_i$
could be allowed to vary at the same time, but this will lead to a degradation in the 
measurement accuracy for all of them \cite{TheLIGOScientific:2016src}; statistical errors will be much smaller when 
the $\delta\hat{p}_i$ are varied one at a time. Note that in most alternative theories of gravity, 
a violation will likely show up in more than one coefficient. However, as already demonstrated in 
\cite{Sampson:2013lpa} in a PN context, looking for a deviation from zero in a single testing 
parameter is an efficient
way to search for GR violations that occur at multiple PN orders, and one can even 
find violations at powers of frequency that are distinct from the one that the testing
parameter is associated with \cite{Li:2011cg,Li:2011vx}. Of course, if a deviation is present then the 
individual measurements of the $\delta\hat{p}_i$ will not necessarily reflect the predicted values
of the correct alternative theory. Should one want to measure or constrain \emph{e.g.}~extra 
charges or coupling constants that may be present in one's favorite alternative theory using
an IMR signal, then an accurate IMR waveform model would need to be constructed for 
that particular theory. However, this is not the aim of the framework presented here; 
what we want to do is test Einstein's theory itself by constraining 
deviations from the theory. 

Even though the IMRPhenomPv2 waveform model has an explicit analytic expression, in the case 
of low-mass binary mergers, which leave a long signal in the detectors' sensitive band, the analyses are
computationally costly and can take more than a month time to complete, due to the large number of 
likelihood evaluations ($\mathcal{O}(10^8)$) that must be performed. Given the large number of 
detections that are expected to be made in the coming LIGO-Virgo observing runs, ways must be found to 
reduce the computational burden. One solution is to speed up the likelihood calculation by constructing 
\emph{reduced-order quadratures} (ROQs) \cite{Antil:2012wf,Canizares:2013ywa,Smith:2016qas}, which in turn 
are based
on reduced-order models 
\cite{Field:2011mf,Field:2013cfa,Purrer:2014fza,Purrer:2015tud,Blackman:2015pia,Blackman:2017dfb,OShaughnessy:2017tak,Blackman:2017pcm}. 
In the method of reduced-order quadratures, 
the discrete overlap calculation involved in the likelihood evaluation is split up into 
a data dependent sum which only needs to be evaluated once for each detection, and 
a much shorter sum that takes care of the parameter dependent part of the overlap calculation 
that must be performed many times during the sampling over parameter space.
In line with the method outlined above, a series of ROQs is created, in each of which
a single testing parameter $\delta\hat{p}_i$ is allowed for.

Results of the parameterized tests for the LIGO-Virgo detections of binary black hole
coalescences have already appeared elsewhere 
\cite{TheLIGOScientific:2016src,TheLIGOScientific:2016pea,Abbott:2017vtc}. 
The aim of this paper is twofold: (a) to construct ROQs for IMRPhenomPv2 waveforms with parameterized
deformations, for use on future detections; and (b) to explicitly demonstrate the robustness and 
sensitivity of the method as a whole, which had not yet been done in previous publications. 

The structure of this paper is as follows. In Sec.~II we briefly recall the 
waveform model used, and explore analytically how the phase varies with the chosen 
deformations, and as a function of mass. We describe the setup of the parameterized tests, 
and explain how
results from multiple detections can be combined to arrive at stronger bounds on GR violations. 
Sec.~III describes the construction of the reduced-order quadrature for waveforms with testing parameters. 
Next, in Sec.~IV we present some checks of the correctness and robustness of the data analysis pipeline. 
In Sec.~V we show how well the parameterized tests can bound GR violations by combining information 
from all available sources. Furthermore, we investigate how testing parameters in the 
template waveforms respond when 
deviations in one or more parameters are present in the signal. Sec.~VI provides a summary 
and conclusions.

Throughout this paper we set $c = G = 1$ unless specified otherwise.

\section{Waveform model and parameterized tests}
\label{sec:waveform}

\subsection{Waveform model}
\label{sec:waveform_model}

The starting point for the parameterized tests is the phenomenological frequency domain 
waveform model which in the LIGO Algorithm Library is designated as IMRPhenomPv2 \cite{Hannam:2013oca}. 
This waveform model describes 
an approximate signal of a precessing binary by applying a rotation transformation 
\cite{Schmidt:2012rh,Hannam:2013oca} to an underlying aligned spin waveform model, 
here taken to be IMRPhenomD \cite{Husa:2015iqa,Khan:2015jqa}. The orbital precession dynamics 
are given in terms of an effective spin parameterisation \cite{Hannam:2013oca,Schmidt:2014iyl}. 
For an in-depth description 
we refer to these papers; here we only give a quick overview.

The phasing of IMRPhenomPv2 consists of three regimes, whose physical meaning and
parameterization are as follows:
\begin{enumerate}
\item The \emph{inspiral regime} is parameterized by post-Newtonian coefficients 
$\{\varphi_0, \ldots, \varphi_7\}$ and $\{\varphi_{5l}, \varphi_{6l}\}$, as well as phenomenological
parameters $\{\sigma_0, \ldots, \sigma_4\}$. The latter are contributions at high effective
PN order (up to 5.5PN) to correct for non-adiabaticity in late inspiral and for unknown
high-order PN coefficients in the adiabatic regime.
\item The \emph{intermediate regime} transitions between inspiral and merger-ringdown; it
is parameterized by the phenomenological coefficients $\{\beta_0, \ldots, \beta_3\}$. 
\item The \emph{merger-ringdown regime}, parameterized by a combination of phenomenological 
and analytical black-hole perturbation theory parameters $\{\alpha_0, \ldots, \alpha_5\}$. 
\end{enumerate}
Note that the PN coefficients $\{\varphi_0, \ldots, \varphi_7\}$ and 
$\{\varphi_{5l}, \varphi_{6l}\}$ have their usual dependences on the binary's
component masses and spins. The other, phenomenological parameters are fixed
by calibration against numerical relativity waveforms. For the functions of frequency in which the above 
parameters appear, we refer to 
\cite{Khan:2015jqa}; see also Table I in \cite{TheLIGOScientific:2016src}. The transition from the inspiral to the intermediate
regime happens at a frequency $f = f_1 = 0.018/M$ (where $M$ is the total mass), and 
from the intermediate to the merger-ringdown regime at $f = f_2 = 0.5\,f_{\rm RD}$, with
$f_{\rm RD}$ a ``ringdown frequency", in such a way that the waveform is $C^1$ continuous. 

\begin{figure}
\includegraphics[width=\columnwidth]{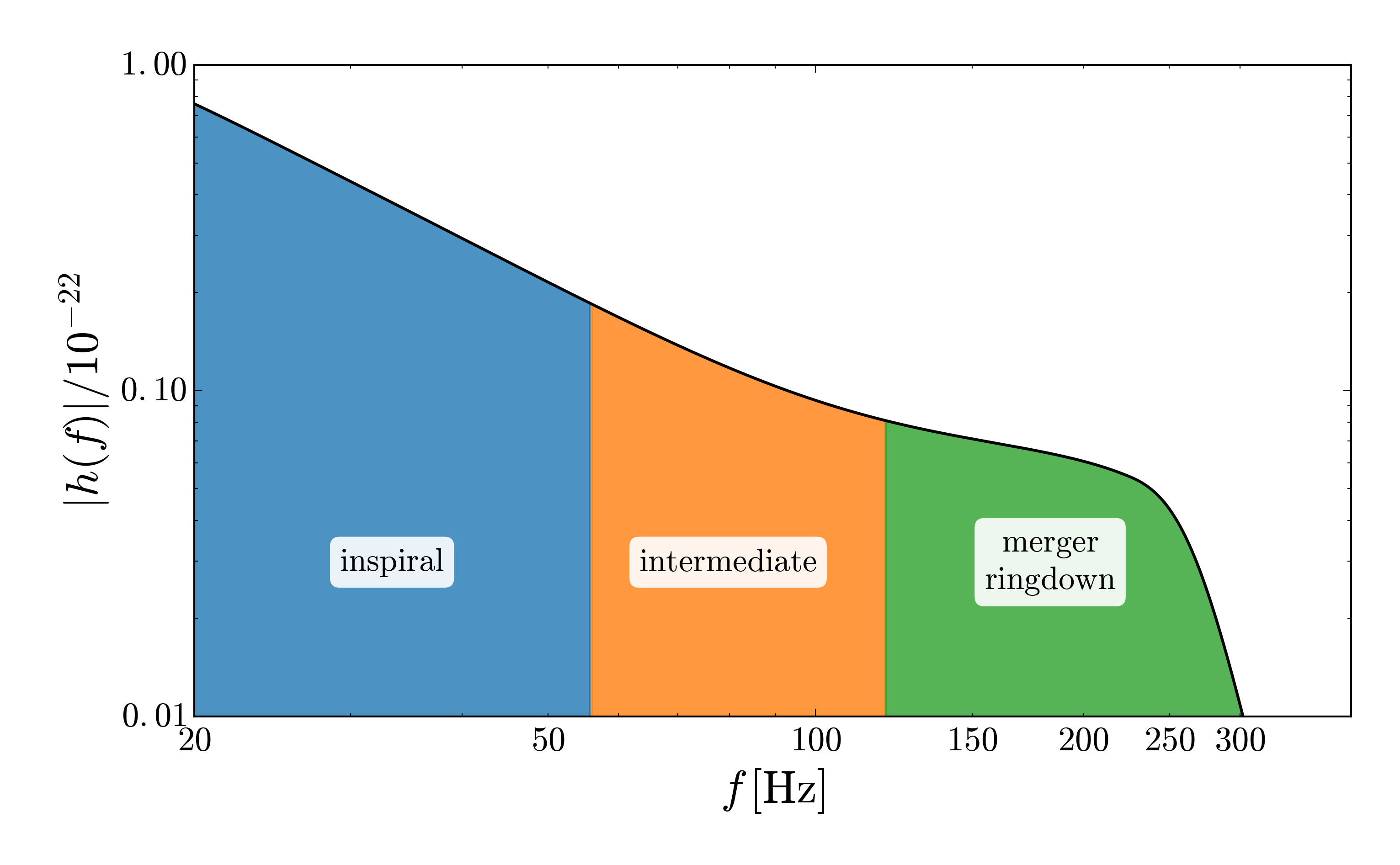}
\includegraphics[width=\columnwidth]{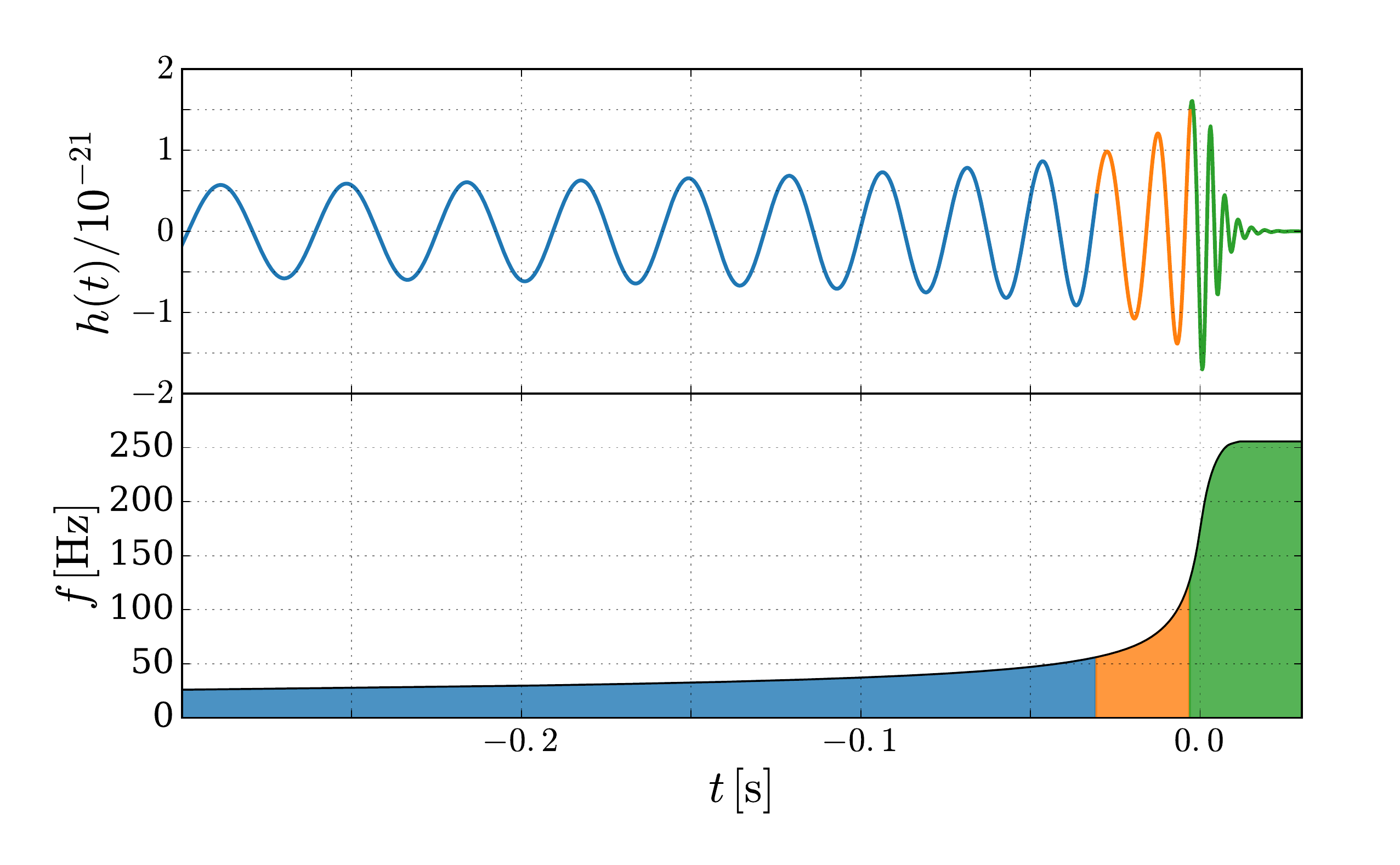}
\caption{The three regimes of the IMRPhenomPv2 model. Top: The modulus of the 
waveform as a function of frequency for a signal similar to GW150914. Bottom: 
Fourier transform to the time domain (top panel) and instantaneous frequency as a 
function of time (bottom panel).}
\label{fig:IMRPhenomPv2}
\end{figure}

Fig.~\ref{fig:IMRPhenomPv2} shows the modulus of the
waveform $|\tilde{h}(f)|$, highlighting the three regimes; also shown is the Fourier 
transform to the time domain, $h(t)$, and the corresponding instantaneous frequency as
a function of time.  

Not all of the coefficients mentioned above will be used in the parameterized tests. 
In the inspiral regime, $\varphi_5$ is completely degenerate with the phase at 
coalescence, $\varphi_c$; similarly, the pair $(\sigma_0, \sigma_1)$ is degenerate
with $(\varphi_c, t_c)$, with $t_c$ the time at coalescence. The pairs $(\beta_0, \beta_1)$
and $(\alpha_0, \alpha_1)$ are set by the requirement of $C^1$ continuity between the 
different regimes. We also omit $\alpha_5$, which occurs in the same term as 
$\alpha_4$, meaning that there will be some amount of degeneracy between the 
two. 
Finally, we do not use $\{\sigma_2, \sigma_3, \sigma_4\}$, whose fractional calibration
uncertainties were larger ($\mbox{(a few)} \times 10^{-1}$)
than those of the other phenomenological parameters (at most $\mbox{(a few)} \times 10^{-2}$),
though all calibration uncertainties were observed to be below measurement uncertainties 
for the binary black hole coalescence detections that were made \cite{PrivateCommunication}. 

The way our parameterized tests are implemented is by allowing for fractional deviations from 
the GR values for all of the remaining coefficients $p_i$ in turn:
\be
p_i^{\rm GR}(m_1, m_2, \mathbf{S}_1, \mathbf{S}_2) \rightarrow 
(1 + \delta\hat{p}_i)\, p_i^{\rm GR}(m_1, m_2, \mathbf{S}_1, \mathbf{S}_2),
\ee
where $m_1$, $m_2$ are the component masses and $\mathbf{S}_1$, $\mathbf{S}_2$ the 
component spins; one has 
\be
\left\{ \delta\hat{p}_i \right\}_i = \{ \delta\hat{\varphi}_0, \ldots, \delta\hat{\varphi}_7, \delta\hat{\varphi}_{5l},
\delta\hat{\varphi}_{6l}, \delta\hat{\beta}_2, \delta\hat{\beta}_3, \delta\hat{\alpha}_2, 
\delta\hat{\alpha}_3, \delta\hat{\alpha}_4 \}.
\ee
We note that in GR, $\varphi_1 \equiv 0$, so that as an exception we let  
$\delta\hat{\varphi}_1$ be an absolute rather than a relative deviation.

Including extrinsic parameters coming from the detector response, in 
practice the full parameter sets of the resulting waveform models will be:
\be
\vec{\lambda} = \{t_c, \varphi_c, D_{\rm L}, \theta, \phi, \psi, m_1, m_2, \chi_1, \chi_2, 
\chi_p, \theta_J, \alpha_0, 
\delta\hat{p}_i\}.
\label{parameters}
\ee
Here $t_c$ and $\varphi_c$ are, respectively, the time and phase at coalescence; 
$D_{\rm L}$ is the luminosity distance; $(\theta, \phi)$ give the sky position; $\psi$ 
is a polarization 
angle; $m_1$ and $m_2$ are the component masses; $\chi_1$ and $\chi_2$ are spin magnitudes; 
and $\chi_p$ is an ``effective" spin precession
parameter given by~\cite{Schmidt:2014iyl}
\be
\chi_p = \frac{\mbox{max} (A_1 m_1^2 \chi_{1 \perp}, A_2 m_2^2 \chi_{2 \perp})}{A_1^2 m_1^2},
\ee
where $A_1 = 2 + 3 m_2/2m_1$, $A_2 = 2 + 3 m_1/2m_2$ and $\chi_{1 \perp}$, $\chi_{2 \perp}$
are the projections of the spin vectors onto the orbital plane, \emph{i.e.}~orthogonal to the 
direction of the orbital angular 
momentum $\hat{L}$ at a specific reference frequency $f_\mathrm{ref}$. 
$\theta_J$ is the angle between 
the line of sight $\hat{n}$ and the total angular momentum $\hat{J}$
at $f_\mathrm{ref}$, and $\alpha_0$ indicates 
the azimuthal 
orientation of $\hat{L}$ at $f_\mathrm{ref}$~\cite{Hannam:2013oca}. 

\subsection{Effect of testing parameters on the phase}

Before going on to evaluate the sensitivity of parameterized tests given 
stellar mass BBHs as seen in the advanced detectors, we first illustrate 
the effect on the phase of varying the $\delta\hat{p}_i$. As it turns out, 
one of the best-determined PN testing parameters tends to be $\delta\hat{\varphi}_3$;
in the intermediate regime this is $\delta\hat{\beta}_2$, and in the merger-ringdown
regime, $\delta\hat{\alpha}_2$; these are the parameters we focus on.

\begin{figure*}
\begin{tabular}{c c}
\includegraphics[width=9cm]{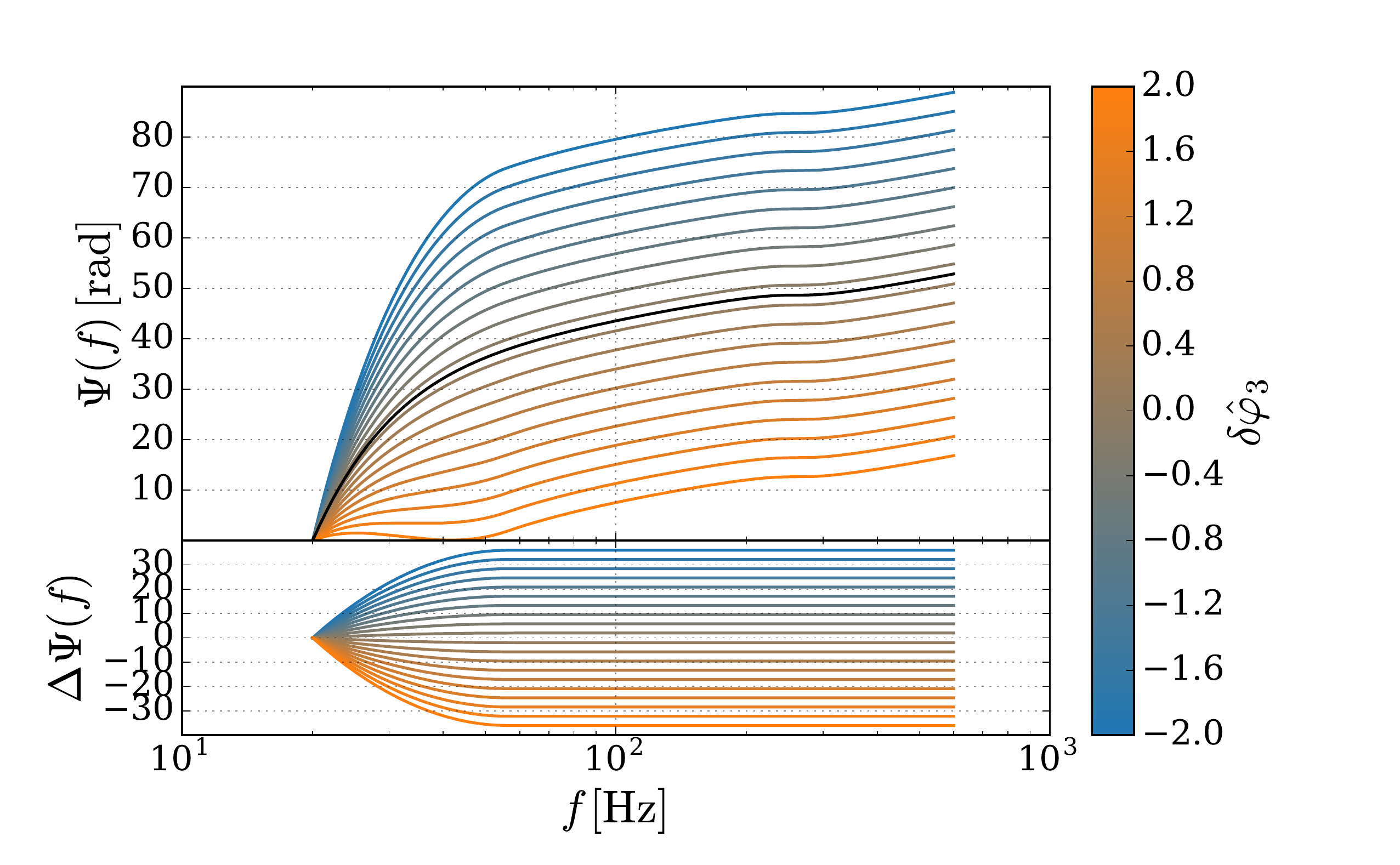} &
\includegraphics[width=9cm]{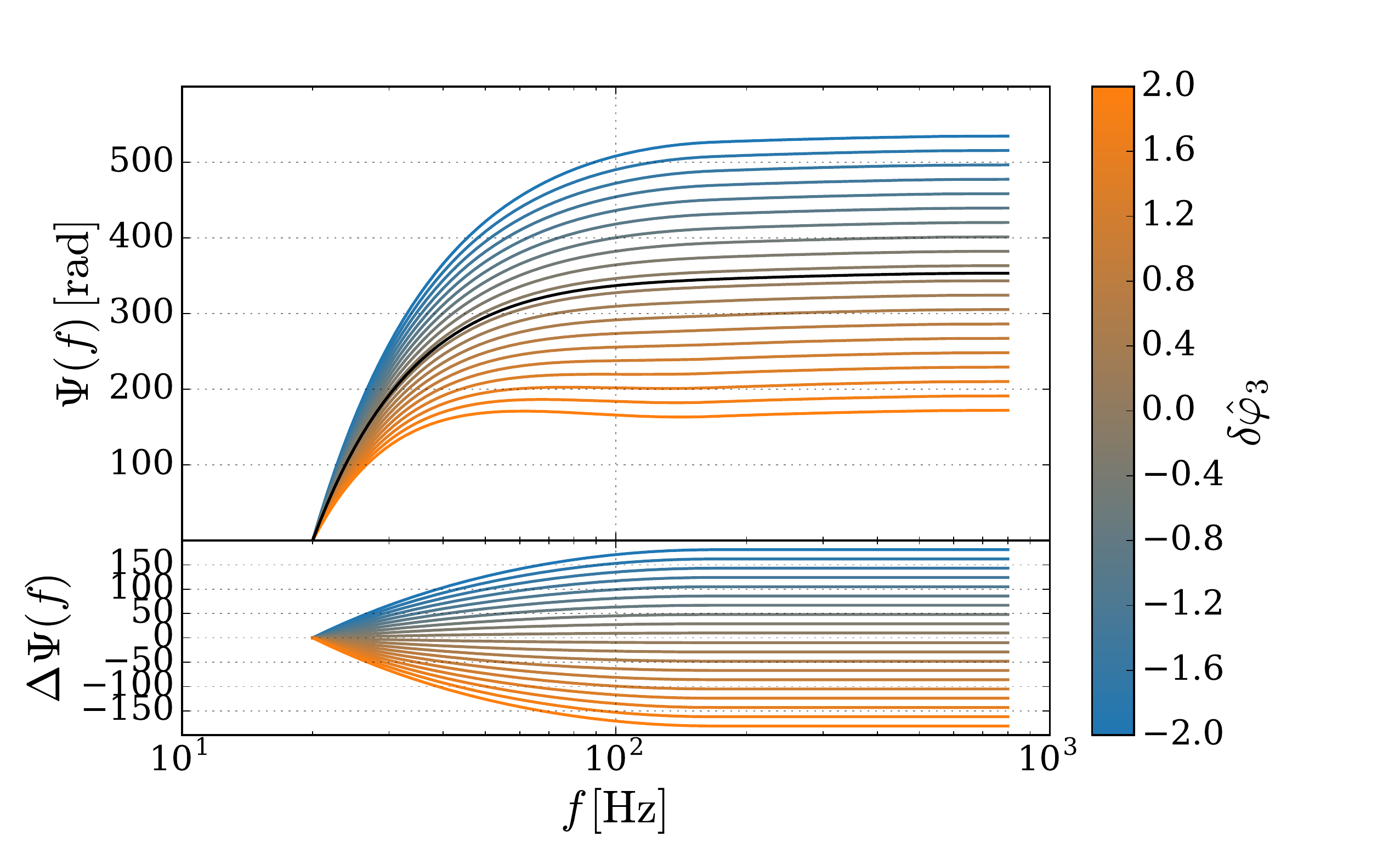} \\
\includegraphics[width=9cm]{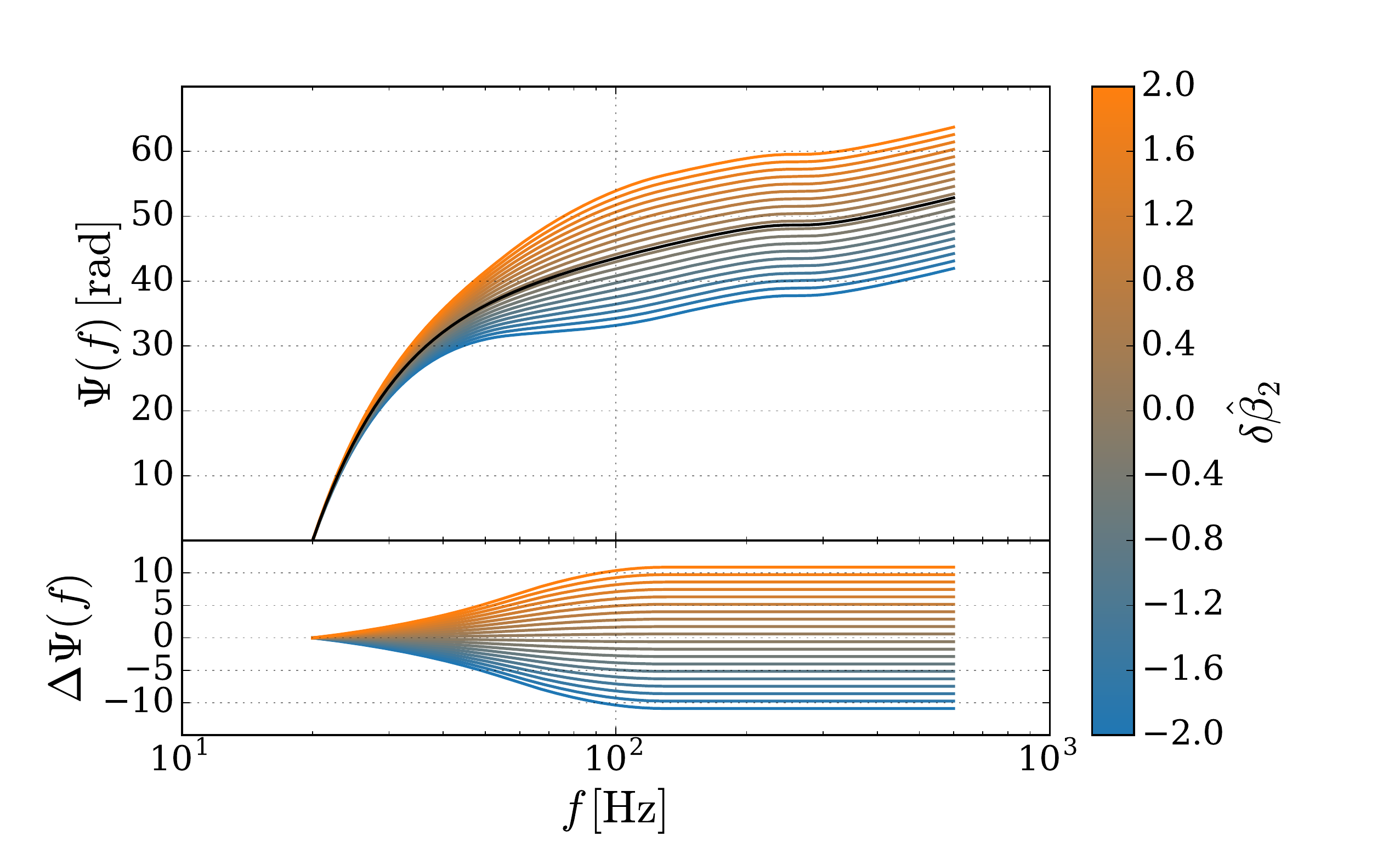} &
\includegraphics[width=9cm]{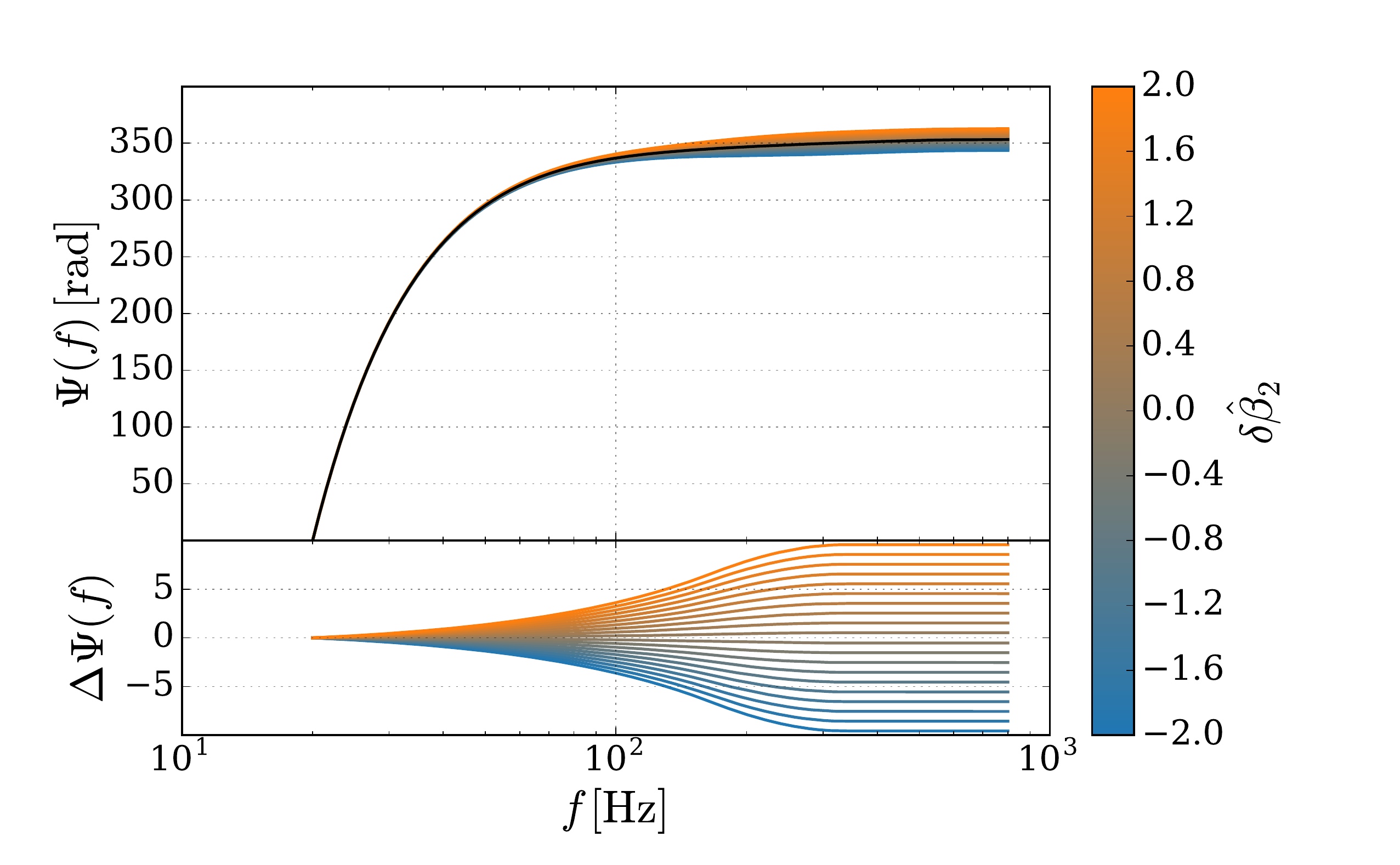} \\
\includegraphics[width=9cm]{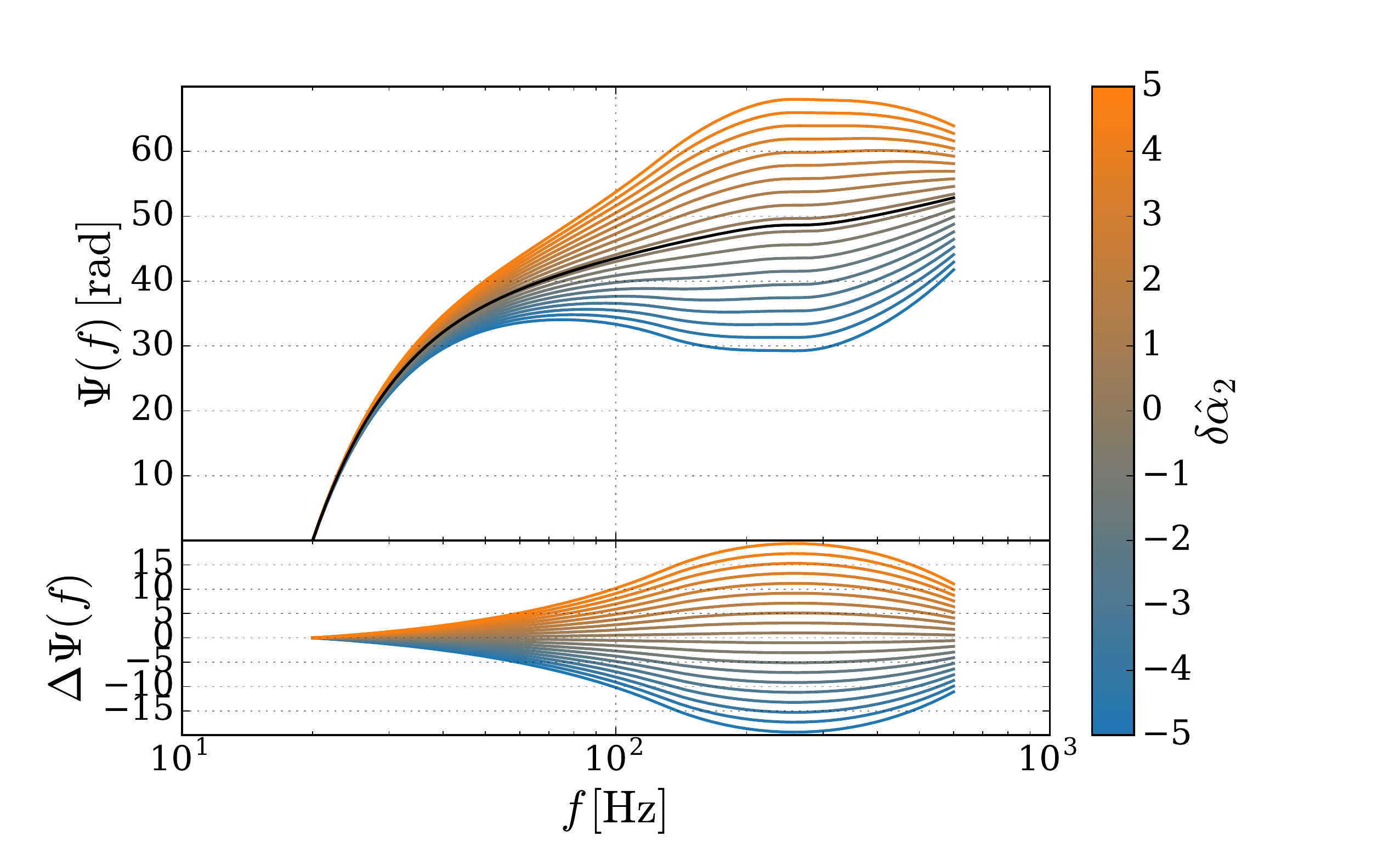} &
\includegraphics[width=9cm]{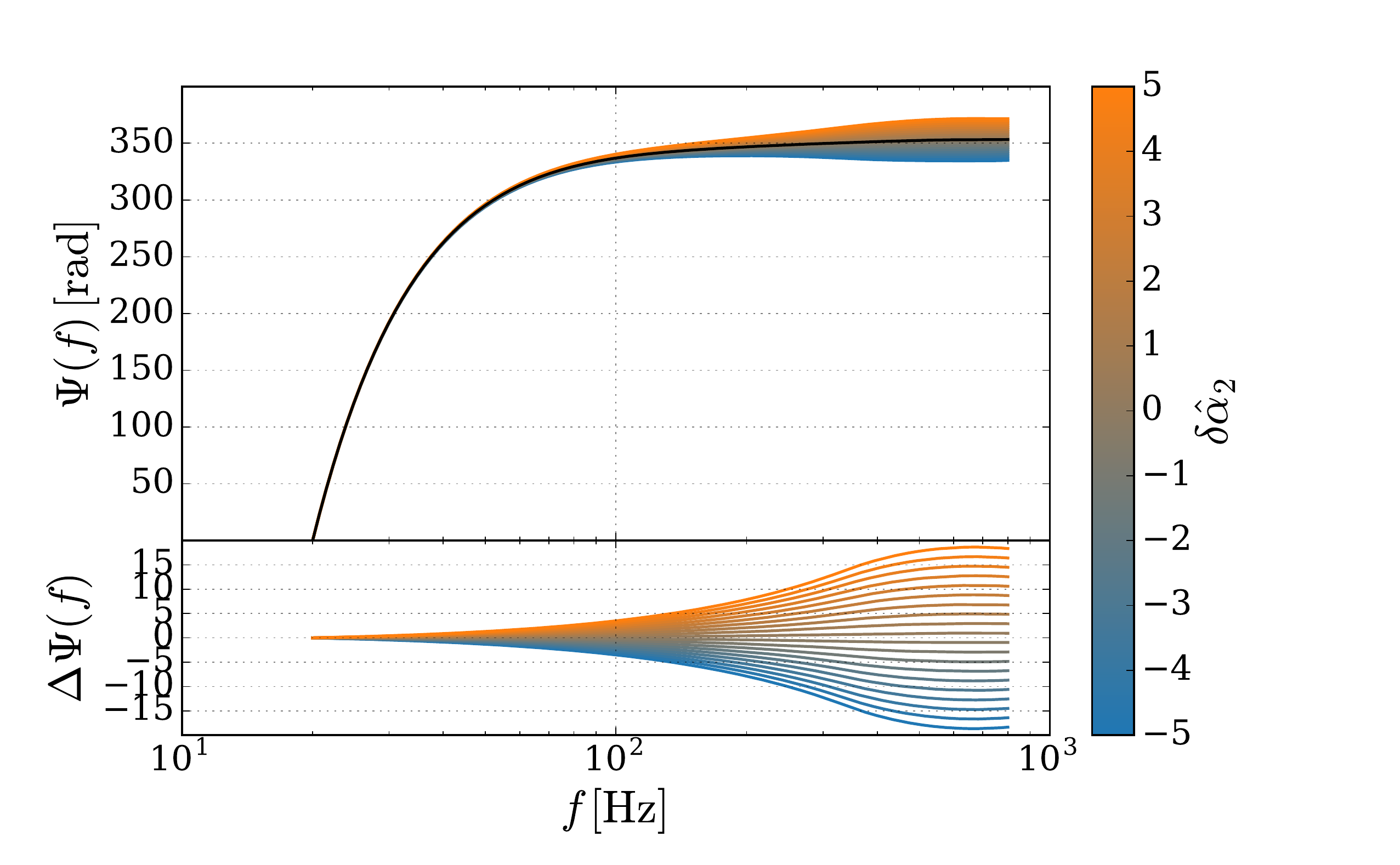} 
\end{tabular}
\caption{The effect of varying testing parameters on the phase as a function of 
frequency, for $t_c = \varphi_c = 0$. In the top of each panel we plot the GR phase (black) as well as 
the way the phase varies with a testing parameter (colors); the bottom 
shows the difference. In the left column we show results for an event with
parameters like that of GW150914, and similarly in the right column for 
GW151226.}
\label{fig:varyingparameters}
\end{figure*}

Fig.~\ref{fig:varyingparameters} shows how the phase as a function of
frequency varies with testing parameters, $\Psi(\delta\hat{p}_i; f)$, 
as well as the difference with the phase in GR, $\Delta\Psi(\delta\hat{p}_i; f) 
= \Psi(\delta\hat{p}_i; f) - \Psi_{\rm GR}(f)$, for $t_c = \varphi_c = 0$. Two kinds of sources 
are considered, with masses and spins 
chosen to be the means of the posterior density functions for the signals that
were designated GW150914 \cite{Abbott:2017xlt} 
and GW151226 \cite{Abbott:2016nmj}. The phases
and their differences are
plotted from $f_{\rm low} = 20$ Hz and up to a frequency where the dominant $(l=2,m=2)$ mode of the
ringdown signal can be safely assumed to have ended (600 Hz for GW150914 and 800 Hz for GW151226). 
The qualitative 
behavior is as expected given the differences between the two. GW150914, being
more massive, had a short inspiral regime, and the merger occurred at $f \sim 130$ Hz, 
close to the frequency where the detectors are the most sensitive. By contrast, GW151226 had
a much longer inspiral (with $\sim 55$ cycles in band), and its merger occurred 
at $f \sim 450$ Hz. For the PN testing parameter
$\delta\hat{\varphi}_3$, a much larger phase difference is accumulated in the case
of GW151226, which will cause this parameter to be much better measured in the latter case 
despite the overall smaller signal-to-noise ratio. 
The intermediate regime parameter $\delta\hat{\beta}_2$ exhibits a relatively slowly 
increasing phase difference and levels out between 100 and 200 Hz for both events, 
which is where the detectors are the most sensitive; hence we can expect it to 
be roughly equally well measurable for both events. Finally, for $\delta\hat{\alpha}_2$,
in the case of GW150914 the phase difference reaches a maximum at some point before 
decreasing again, while for GW151226 the difference keeps increasing up to high frequencies, 
but not with a larger maximum phase difference; hence this parameter will be
better measurable with GW150914, for which the merger-ringdown regime occurs at frequencies
closer to the range of best detector sensitivity. These expectations are borne out
by the published results for the two events 
\cite{TheLIGOScientific:2016src,TheLIGOScientific:2016pea,Abbott:2017vtc}. 
Note that varying the $\delta\hat{p}_i$ has an effect at \emph{all} frequencies; this
is a consequence of the $C^1$ junction conditions between the inspiral, intermediate, 
and merger-ringdown regimes. 

\begin{figure}
\includegraphics[width=9cm]{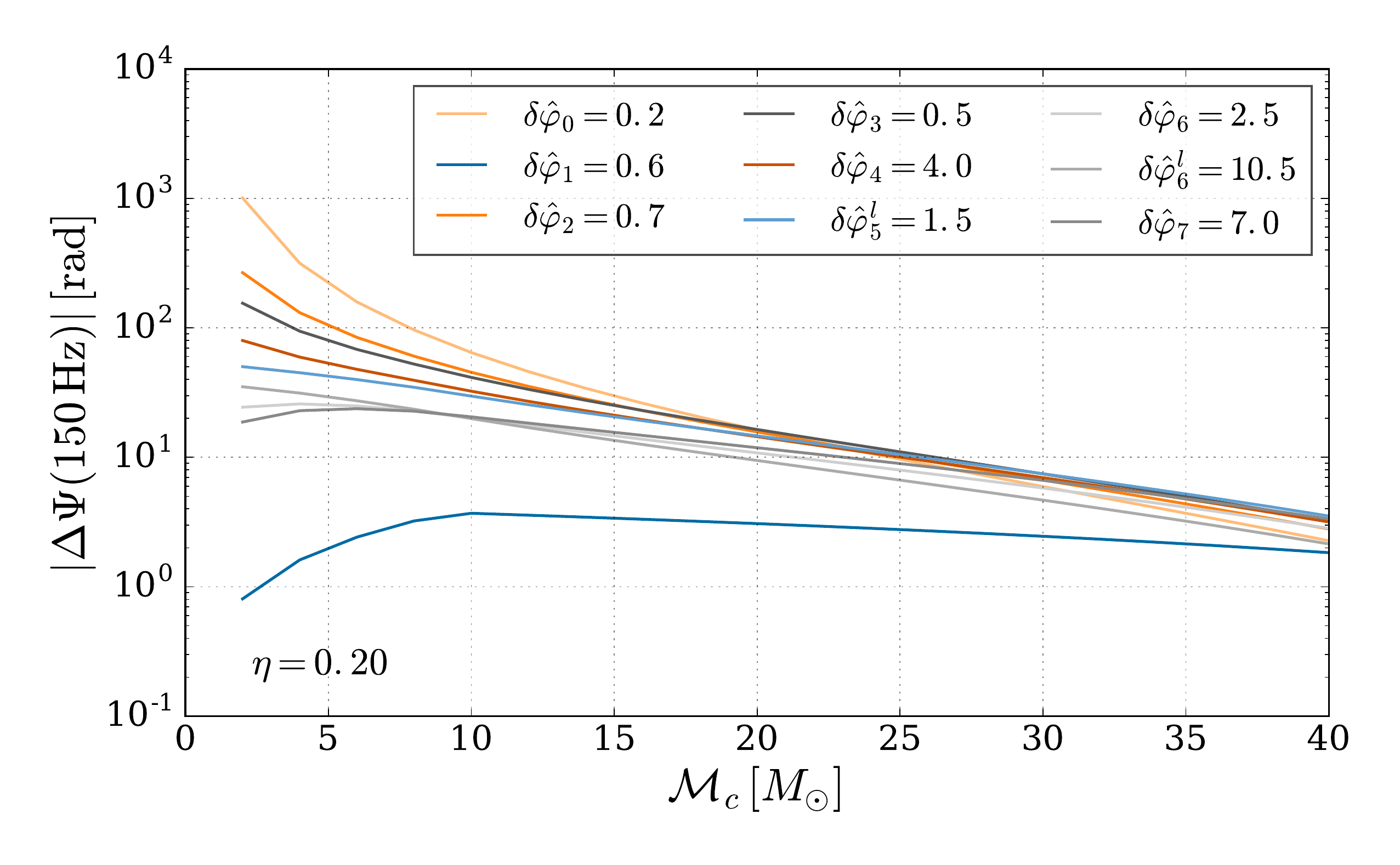} 
\includegraphics[width=9cm]{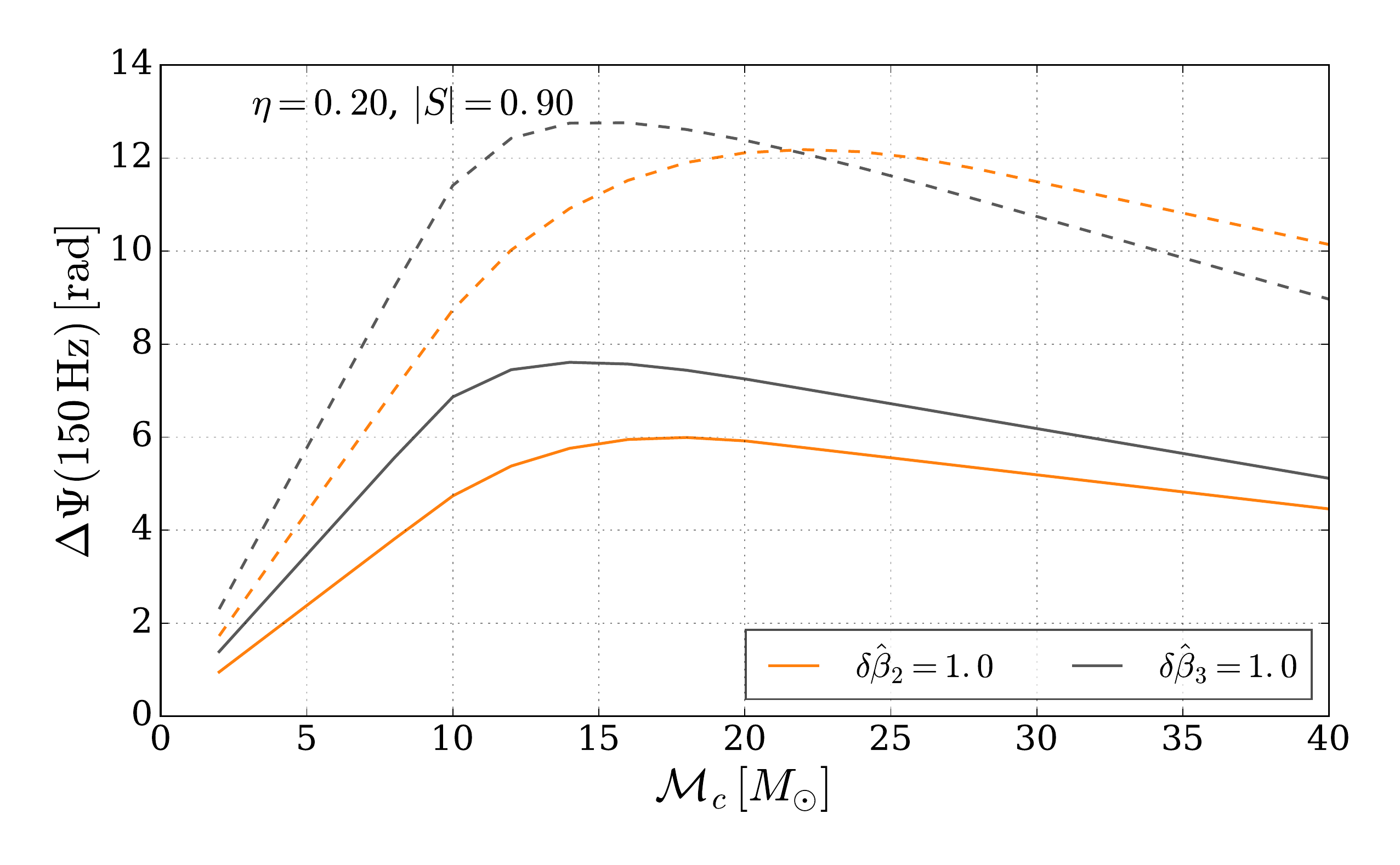} 
\includegraphics[width=9cm]{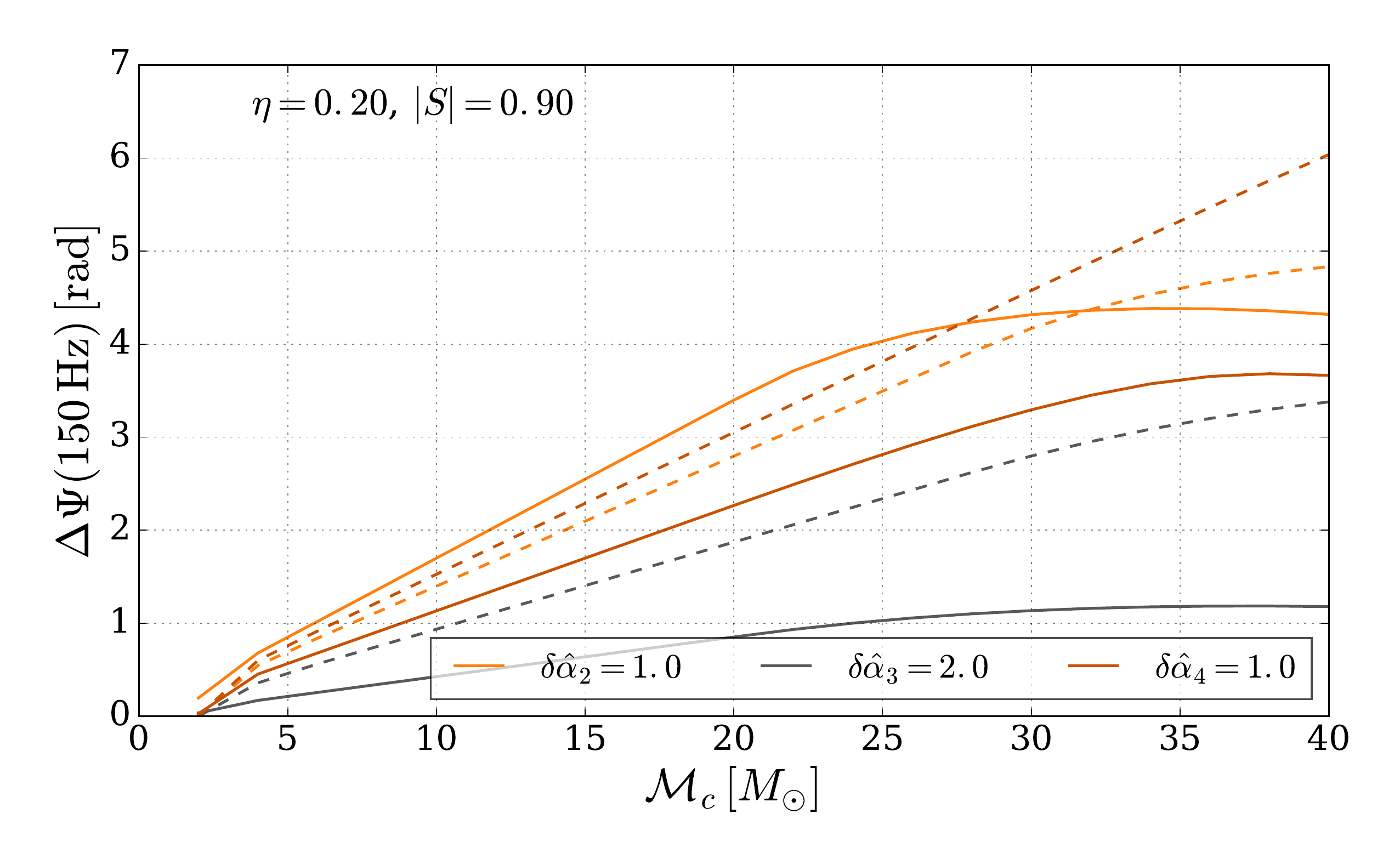}
\caption{The differences at $f = 150$ Hz between the GR phase and the phase for 
given values of testing parameters, for inspiral (top), the intermediate regime (middle),
and the merger-ringdown regime (bottom). For definiteness we again set $t_c = \varphi_c = 0$. 
The solid lines are for zero spins, and the dashed lines
for aligned spins with $|\mathbf{S}_{1,2}| = 0.9$.}
\label{fig:varyingmchirp}
\end{figure}

In Fig.~\ref{fig:varyingmchirp} we illustrate the phase differences 
as chirp mass $\mathcal{M}_c = M \eta^{3/5}$ is varied (where $M = m_1+m_2$ and 
$\eta = m_1 m_2/M^2$), for particular values of the $\delta\hat{p}_i$; the 
symmetric mass ratio is fixed at $\eta = 0.2$, and again $t_c = \varphi_c = 0$; shown are the 
$\Delta\Psi(\delta\hat{p}_i; f)$
for $f = 150$ Hz, \emph{i.e.}~close to the frequency of optimal sensitivity for the
Advanced LIGO detectors. Again the behavior is as expected. Low $\mathcal{M}_c$ 
corresponds to waveforms with significant inspiral in band; deviations in the $\varphi_i$
then have a large effect on the observable phase. Deviations in the 
intermediate regime parameters $\beta_i$
have the largest effect when this regime occurs at frequencies 
where the detectors are the most sensitive, which corresponds to
$\mathcal{M}_c = 10 - 20\, M_\odot$. Finally, shifts in the merger-ringdown parameters
$\alpha_i$ have their largest effect for $\mathcal{M}_c \gtrsim 20 
M_\odot$, which brings this 
regime in the detectors' most sensitive band.  
For completeness we show results 
for zero spins, as well as aligned spins with $|\mathbf{S}_{1,2}| = 0.9$; the well-known 
effect of ``orbital hang-up" \cite{Campanelli:2006fy}, which prolongs the duration of waveforms in the time 
domain, then causes similar features to occur at higher $\mathcal{M}_c$.

\subsection{Parameter estimation}

Parameter estimation is done using the LALInference framework \cite{Veitch:2009hd,Veitch:2014wba}, in which
the posterior density distribution for the parameters $\vec{\lambda}$ is obtained as
\be
p(\vec{\lambda} | H_i, d, I) = \frac{p(\vec{\lambda} | H_i, I)\,p(d | H_i, \vec{\lambda}, I)}{p(d|I)}.
\ee
Here $H_i$ is the hypothesis corresponding to the waveform model in which $\delta\hat{p}_i$
is an extra free parameter, $d$ is the data, and $I$ denotes whathever background information
we may have. $p(d | H_i, \vec{\lambda}, I)$ is the likelihood function, which up to 
an overall prefactor is given by
\be
p(d | H_i, \vec{\lambda}, I) \propto 
\exp[-\langle d - h(\vec{\lambda}) | d - h(\vec{\lambda}) \rangle/2],
\label{likelihood}
\ee
with $h(\vec{\lambda})$ the signal model described in Sec.~\ref{sec:waveform_model}, 
and $\langle \, \cdot \,|\, \cdot \,\rangle$ the noise-weighted inner product
\be
\langle a | b \rangle  
= 4 \Re \int_{f_{\rm low}}^{f_{\rm high}} \frac{a^\ast(f) b(f)}{S_n(f)} \,df,
\ee
where $S_n(f)$ is the one-sided noise power spectral 
density. For the second-generation detectors, the lower cut-off frequency is taken 
to be $f_{\rm low} = 20$ Hz, while $f_{\rm high} = 2048$ Hz suffices as an upper cut-off
frequency for stellar mass BBH coalescences. The likelihood function is evaluated using
the efficient Nested Sampling algorithm \cite{Veitch:2009hd}. $p(\vec{\lambda} | H_i, I)$ is the 
prior probability density for the free parameters; for those parameters that also appear in the 
GR waveform these are chosen in the same way as in \cite{Veitch:2014wba}, while for $\delta\hat{p}_i$ 
we choose priors uniform in an interval that is wide enough to contain the supports of the posterior 
densities; suitable ranges are given in subsection \ref{subsec:IMRPhenomPv2_ROQ}
below. Lastly, $p(d|I)$ is the probability of the data, which can be absorbed
into an overall normalization factor for the posterior density $p(\vec{\lambda} | H_i, d, I)$.

To obtain one-dimensional posterior densities for the parameters $\delta\hat{p}_i$, one
marginalizes over all other parameters:
\be
p(\delta\hat{p}_i | H_i, d, I) 
= \int d\vec{\theta}\,p(\vec{\theta}, \delta\hat{p}_i | H_i, d, I), 
\ee
where the integration is performed over all parameters in (\ref{parameters}) except for $\delta\hat{p}_i$.

Finally, posterior densities from individual events can be conveniently combined to arrive at 
stronger bounds on the $\delta\hat{p}_i$ under the assumption that the fractional deviations 
are the same in each event. 
Assuming independent detections $d_1, d_2, \ldots, d_N$, it is easy to see that
\ba
&& p(\delta\hat{p}_i | H_i, d_1, d_2, \ldots, d_N, I) \nonumber\\
&& = p(\delta\hat{p}_i|I)^{1-N} \prod_{n=1}^N p(\delta\hat{p}_i | H_i, d_n, I).
\label{combinedposteriors}
\ea 
For events with similar signal-to-noise ratios and in the absence of measurement offsets, 
one can expect the widths of these posteriors to decrease roughly with $\sqrt{N}$.

\section{Reduced-order quadratures for fast likelihood calculations}
\label{sec:ROQ}

\subsection{Basic method}

We now proceed to constructing reduced-order quadratures. The technical underpinnings have 
already been discussed in detail elsewhere 
\cite{Canizares:2013ywa,Smith:2016qas}; here we will only give an overview. 

The first 
step is to approximate the waveform $h(\vec{\lambda}; f)$ as 
\ba
h(\vec{\lambda}; f) &\simeq& \mathcal{P}_{\mathcal{E}_n}[h(\vec{\lambda}; f)] \nn\\
&\equiv& \sum_{i=1}^n ( e_i | h(\vec{\lambda}) )\,e_i(f),
\ea
where the vectors in the \emph{reduced basis} $\mathcal{E}_n = \{e_i(f)\}_{i=1}^n$ are orthonormal 
with respect to the inner product 
\be
(a | b) \equiv \int_{f_{\rm min}}^{f_{\rm max}} a^\ast(f)\,b(f)\,df,
\ee
and the approximation is good to within a \emph{greedy projection error} $\epsilon$:
\be
|| h(\vec{\lambda}; f) - \mathcal{P}_{\mathcal{E}_n}[h(\vec{\lambda}; f)] ||^2 < \epsilon \, ,
\label{tolerancecriterion}
\ee
for $\vec{\lambda} \in \mathcal{T}_N$, where $\mathcal{T}_N$ is a suitably large \emph{training set}, 
and $||a|| \equiv \sqrt{\left( a | a \right)}$.
From this, one constructs an \emph{empirical interpolant} to approximate the waveform: 
\be
\mathcal{I}_n[h](\vec{\lambda}; f) \equiv \sum_{i=1}^n x_i(\vec{\lambda})\,e_i(f),
\ee
where the coefficients $x_i$ are solutions to
\be
\mathcal{I}_n[h](\vec{\lambda}; \mathcal{F}_k) = h(\vec{\lambda}; \mathcal{F}_k)
\ee
at interpolation points $\{\mathcal{F}_k\}_{k = 1}^n$. Defining the matrix
$A_{ij} = e_j(\mathcal{F}_i)$, one has 
\ba
\mathcal{I}_n[h](\vec{\lambda}; \mathcal{F}_k) 
&=& \sum_{i=1}^n \sum_{k=1}^n (A^{-1})_{ik} h(\vec{\lambda}; \mathcal{F}_k)\,e_i(f) \nn\\
&=& \sum_{k=1}^n B^L_k(f)\,h(\vec{\lambda}; \mathcal{F}_k),
\ea
where 
\be
B^L_k(f) = \sum_{i=1}^n (A^{-1})_{ik} e_i(f).
\ee
The $\{\mathcal{F}\MA{_{k}}\}_{k = 1}^n$ are chosen from a set $\{f _i\}_{i=1}^L$, where 
$L$ is related to the duration $T$ of the longest waveform considered through
\be
L = (f_{\rm max} - f_{\rm min})\,T + 1,
\ee
and the $f_i$ are spaced by $\Delta f = 1/T$.
The first interpolation point $\mathcal{F}_1$ is chosen such that it maximizes the 
amplitude of the first reduced basis vector, \emph{i.e.}~$|e_1(\mathcal{F}_1)| \geq |e_1(f_i)|$
for all $f_i$. Next one builds an interpolant of $e_2(f)$ using only $e_1$ and $\mathcal{F}_1$,
and one finds an $\mathcal{F}_2$ that maximizes the pointwise interpolation error, 
\emph{i.e.}~$|\mathcal{I}_1[e_2](\mathcal{F}_2) - e_2(\mathcal{F}_2)| 
\geq |\mathcal{I}_1[e_2](f_i) - e_2(f_i)|$ for all $f_i$. One then continues in this fashion this until
$n$ interpolation points have been obtained. 

Though the interpolant $\mathcal{I}_n[h](\vec{\lambda}; f)$ can be evaluated at any
parameter values $\vec{\lambda}$, the underlying reduced basis $\mathcal{E}_n$ satisfies the
tolerance criterion (\ref{tolerancecriterion}) only for $\vec{\lambda} \in \mathcal{T}_N$. 
Next comes the \emph{validation step}, where the accuracy of the interpolant is evaluated
also for values $\vec{\lambda}$ that lie outside the training set (though inside 
the same ranges as for the training set, where the waveform approximant is deemed valid). 
Arbitrary values are 
picked, for which it is checked that 
\be
|| h(\vec{\lambda}; f) - \mathcal{I}_n[h](\vec{\lambda}; f) ||^2 < \beta
\ee
for some choice of maximum \emph{interpolation error} $\beta$. All ``bad points" $\vec{\lambda}$ for which this is not the case get 
collected and added to the training set $\mathcal{T}_N$, thus creating a new training set
on which the algorithm is repeated, leading to a new interpolant $\mathcal{I}_{n'}[h](\vec{\lambda}; f)$.
The validation step is repeated until no more ``bad points" are found, leading to the 
final interpolant $\mathcal{I}_{N_L}[h](\vec{\lambda}; f)$.

Recall that the aim is to speed up the calculation of the likelihood 
$\LL = p(d|H_i, \vec{\lambda}, I)$, the logarithm of which takes the form
\be
\log \LL =  
\frac{1}{2} \left[ 
2 \langle d | h(\vec{\lambda}) \rangle  
- \langle h(\vec{\lambda}) | h(\vec{\lambda}) \rangle 
- \langle d|d \rangle
\right].
\label{expandedlikelihood}
\ee
First consider the term $\langle d | h \rangle$. 
Substituting for $h(\vec{\lambda}; f)$ the 
empirical interpolant $\mathcal{I}_{N_L}[h](\vec{\lambda}; f)$ and discretizing the integral
in the definition of the inner product, one gets
\ba
\langle d | h(\vec{\lambda}) \rangle
&=& 4 \Delta f \: \Re \sum_{i=1}^L \frac{d^\ast(f_i)\,h(\vec{\lambda}; f_i)}{S_n(f_i)} \nn\\
&\simeq& 4 \Delta f \: \Re \sum_{i=1}^L \sum_{k=1}^{N_L} B^L_k(f_i)\,h(\vec{\lambda}; \mathcal{F}_k) \frac{d^\ast(f_i)}{S_n(f_i)} \nn\\
&=& 4 \Delta f \: \Re \sum_{k=1}^{N_L} \left[ \sum_{i=1}^L B^L_k(f_i) \frac{d^\ast(f_i)}{S_n(f_i)} \right]\,h(\vec{\lambda}; \mathcal{F}_k) \nn\\
&=& \sum_{k=1}^{N_L} w_k h(\vec{\lambda}; \mathcal{F}_k),
\label{dh}
\ea
where 
\be
w_k \equiv 4 \Delta f \: \Re \sum_{i=1}^L B^L_k(f_i) \frac{d^\ast(f_i)}{S_n(f_i)}.
\ee
An important point is now that typically $N_L \ll L$, and the likelihood 
calculations (\ref{dh}) which during the nested sampling process 
need to be performed ``on the fly" for many different parameter values
now only involve the evaluation of $N_L$ expressions $h(\vec{\lambda}; \mathcal{F}_k)$,
rather than the $L$ evaluations of $h(\vec{\lambda}; f)$ that were required originally.
While it is true that the calculation of the ROQ weights $w_k$ still involves a
sum over $L$ terms, they only need to be evaluated once for every detection. This 
means that with the ROQ, this part of the likelihood calculation will be sped up by a 
factor $L/N_L$.

Finally, in Eq.~(\ref{expandedlikelihood}) there is also the term 
$\langle h(\vec{\lambda}) | h(\vec{\lambda}) \rangle$, which can be approximated by
an expression of the form
\be
\langle h(\vec{\lambda}) | h(\vec{\lambda}) \rangle 
= \sum_{j=1}^{N_Q} w^Q_j |h(\vec{\lambda}; \mathcal{F}^Q_j)|^2,
\ee
where
\be
w^Q_j = 4 \Delta f \Re \sum_{i=1}^L \frac{B^Q_j(f_i)}{S_n(f_i)},
\ee
for some $B^Q_j$, and typically $N_Q \ll L$. The $B^Q_j$
are obtained through a similar procedure as in the linear case. 
Here too, the weights $w^{Q}_{j}$ will only have to be calculated once per detection.
Note that `$L$' and `$Q$' in the superscripts respectively refer to the linear 
and quadratic parts of the likelihood.

The above only gives an overview of the rationale behind ROQs. 
In practice one needs to write the waveform $h(\vec{\lambda}; f)$ 
in terms of the $+$ and $\times$ polarizations as 
$F_+ h_+(\vec{\lambda}; f) + F_\times h_\times(\vec{\lambda}; f)$, with $F_+$ and
$F_\times$ the beam pattern functions. However, it turns out that  
a single set of functions $\{B^L_j\}$ is sufficient to represent $h_+$ and $h_\times$,
and a single set $\{B^Q_j\}$ to represent the products 
$|h_+|^2$, $|h_\times|^2$, and $\Re h^\ast_+ h_\times$ \cite{Canizares:2013ywa,Smith:2016qas}.

\subsection{An ROQ for IMRPhenomPv2 with parameterized deformations}
\label{subsec:IMRPhenomPv2_ROQ}

\begin{figure}
\includegraphics[width=9cm]{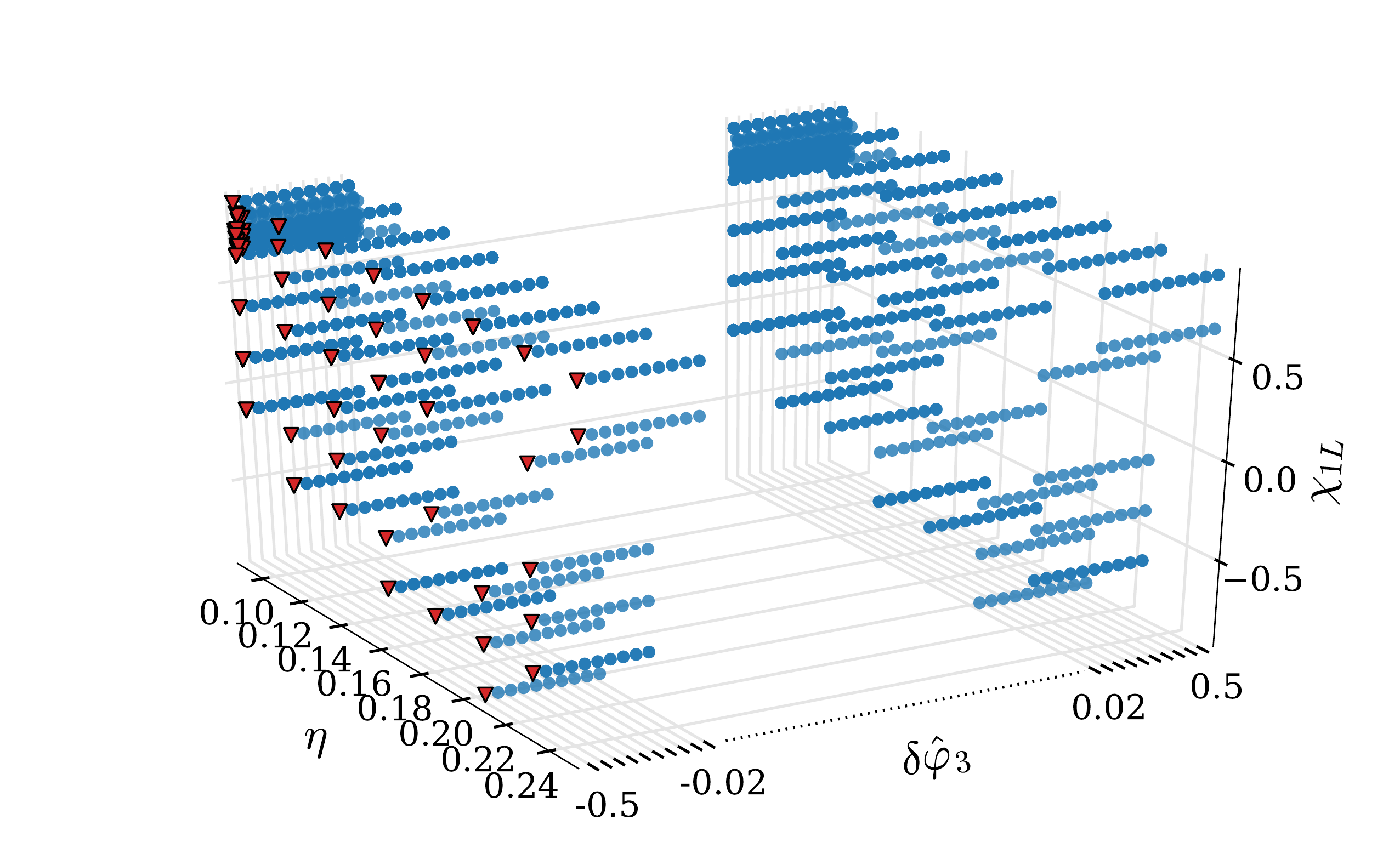} 
\caption{Schematic illustration of how the original basis for IMRPhenomPv2 from 
\cite{Smith:2016qas} (triangles) is extended in the additional parameter dimension $\delta \hat{p}_i$ 
(in this example $\delta\hat{\varphi}_3$) to form a new training set. The plot only shows 
a 3-dimensional slice of the full parameter space. Note that points with 
$-0.02 \leq \delta\hat{\varphi}_3 \leq 0.02$ are not shown to aid visualization.}
\label{fig:basisextension}
\end{figure}

An ROQ for IMRPhenomPv2 in the GR case was already constructed in \cite{Smith:2016qas}. Here we 
want to build a series of ROQs for IMRPhenomPv2, each including a single testing parameter
$\delta \hat{p}_i$. 
As a starting point we use the final reduced basis for the GR waveform, $\mathcal{T}_N$ (where
$N$ can be either the $N_L$ or the $N_Q$ of the linear and quadratic bases, respectively), 
and for each basis element we introduce $N_{\delta \hat{p}_i} = 500$ samples placed uniformly 
in the $\delta \hat{p}_i$ direction; see Fig.~\ref{fig:basisextension}. The ranges for the
various $\delta\hat{p}_i$ are chosen such that they accommodate the widths of posterior density
functions of the LIGO-Virgo events that were recorded so far (with the exception of 
$\delta\hat{\alpha}_2$, $\delta\hat{\alpha}_3$, $\delta\hat{\alpha}_4$, which are 
essentially unmeasurable for low-mass events):
\ba
&&\delta\hat{\varphi}_0 \in [-2, 2],\,\,\,\,\,\,\,\delta\hat{\varphi}_1 \in [-5, 5],\,\,\,\,\,\,\,\delta\hat{\varphi}_2 \in [-10,10], \nn\\
&&\delta\hat{\varphi}_3 \in [-10, 10],\,\,\,\,\,\,\,\delta\hat{\varphi}_4 \in [-10, 10],\,\,\,\,\,\,\,\delta\hat{\varphi}_{5l} \in [-10,10], \nn\\
&&\delta\hat{\varphi}_6 \in [-10, 10],\,\,\,\,\,\,\,\delta\hat{\varphi}_{6l} \in [-20, 20],\,\,\,\,\,\,\,\delta\hat{\varphi}_7 \in [-20,20], \nn\\
&&\delta\hat{\beta}_2 \in [-5, 5],\,\,\,\,\,\,\,\delta\hat{\beta}_3 \in [-5, 5], \nn\\
&&\delta\hat{\alpha}_2 \in [-5, 5],\,\,\,\,\,\,\,\delta\hat{\alpha}_3 \in [-5, 5],\,\,\,\,\,\,\,\delta\hat{\alpha}_4 \in [-5,5].
\ea

\begin{table}
\begin{tabular}{cccccccc}
\hline
Bin & \vline & $\mathcal{M}_c$ ($M_\odot$) & \vline & GR waveform duration (sec) & \vline & $\Delta f$ (Hz) \\
\hline
\hline
A & \vline & [12.3, 45] & \vline & [0.4, 4] & \vline & 1/4 \\
B & \vline & [7.9, 14.8] & \vline & [3, 8] & \vline & 1/8 \\
C & \vline & [5.2, 9.5] & \vline & [6, 16] & \vline & 1/16 \\
D & \vline & [3.4, 6.2] & \vline & [12, 32] & \vline & 1/32 \\
E & \vline & [2.2, 4.2] & \vline & [23.8, 64] & \vline & 1/64\\
\hline
\label{table:chirpmassbins}
\end{tabular}
\caption{The different chirp mass bins for which ROQs were built, with the ranges of waveform durations
in the GR case as well as sampling in frequency.}
\end{table}

The resulting
set $\mathcal{T}_{N \times N_{\delta\hat{p}_i}}$ then becomes a training set for the construction
of a new ROQ, as outlined in the previous subsection. As for the ROQ of the GR waveform, 
this is done independently for waveforms in different, overlapping chirp mass bins, so as to obtain
better likelihood calculation speed-ups than when all chirp masses would be lumped together. 
The chirp mass ranges roughly 
corresponding to different ranges for the length of the waveform in the time domain. 
We note that away from the GR case there is no clear mapping from chirp mass 
to waveform length, as the latter is also partially determined by the values of the $\delta\hat{p}_i$. 
Even so, in each bin we \emph{a priori} set $\Delta f = 1/T_{\rm max}$, where $T_{\rm max}$ is 
the longest GR waveform in the bin; though waveforms can become longer when $\delta\hat{p}_i \neq 0$,
in the end what counts is that all interpolation errors are below the given threshold. 
To reduce the burden on computer memory required, we perform multi-banding as explained 
in \cite{Smith:2016qas}: an adaptive frequency resolution $\Delta f(f)$ is applied such that
waveforms are sampled less densely at higher frequencies, where there is less power  
per frequency bin due to the faster frequency sweep (see Fig.~\ref{fig:IMRPhenomPv2}). However, 
once a basis has been obtained, we up-sample by direct evaluation of the waveform model.

The various bins 
are shown in Table I. Note that no ROQs were made for systems 
with $\mathcal{M}_c > 45\,M_\odot$, since for such binaries the signal will be 
short enough that parameter estimation is sufficiently fast, and not much speed-up
can be expected from an ROQ. The bin with
the lowest chirp masses considered here is $\mathcal{M}_c \in [2.2, 5.2]\,M_\odot$, 
corresponding to a lowest \emph{total} mass of $M \simeq 5\,M_\odot$ for $m_1/m_2 = 1$,
which should suffice for the lightest astrophysical binary black holes. For the other parameters appearing in 
IMRPhenomPv2, we use the same ranges as in \cite{Smith:2016qas}: $1 \leq m_1/m_2 \leq 9$; 
$(-0.9, -0.9, 0) \leq (\chi_{1L}, \chi_{2L}, \chi_p) \leq (0.9, 0.9, 0.9)$, where $\chi_{1L}$,
$\chi_{2L}$ are the spin components along the direction of angular momentum $\hat{L}$; 
$(0,0) \leq (\theta_J, \alpha_0) \leq (\pi, 2\pi)$; and $m_1 \geq m_2 \geq 1\,M_\odot$. 
In the validation steps, we also impose the bound $\chi_{1L} \geq 0.4 - 7\eta$, as was
done in \cite{Smith:2016qas}; this is needed to avoid clustering of bad points 
in a particular region, indicating a limitation of the original IMRPhenomPv2 waveform model. For the ROQs
with the $\delta\hat{p}_i$, it turned out to be necessary to impose an additional 
bound $\sqrt{\chi_{1L}^2 + \chi_p^2} \leq 0.98$. 

\begin{figure*}
\includegraphics[width=8cm]{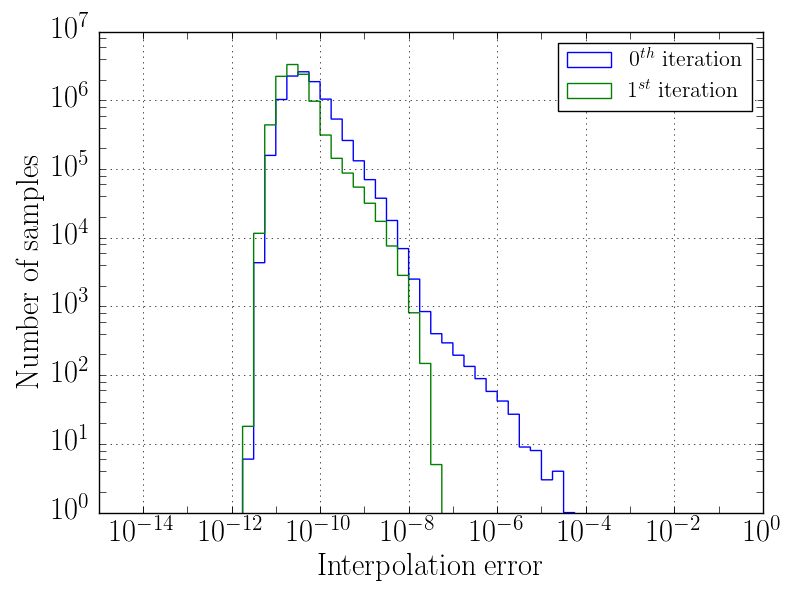}
\includegraphics[width=8cm]{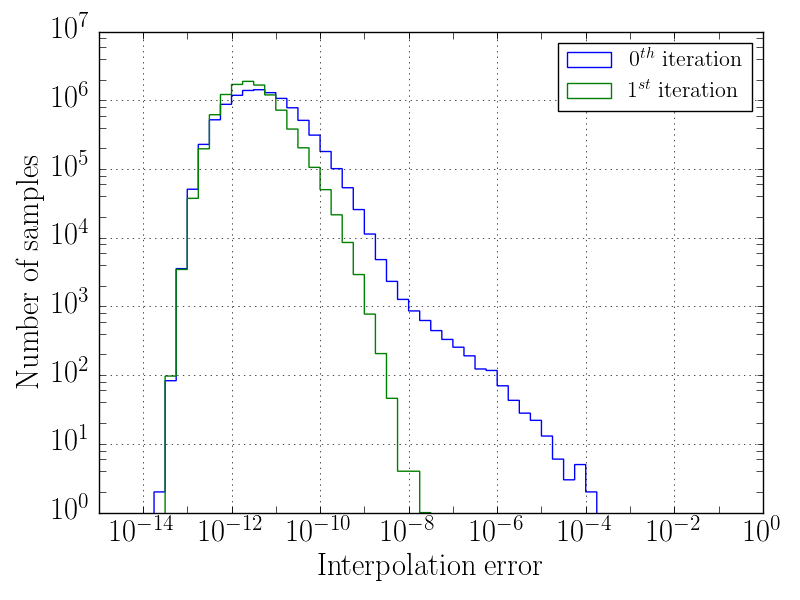}
\includegraphics[width=8cm]{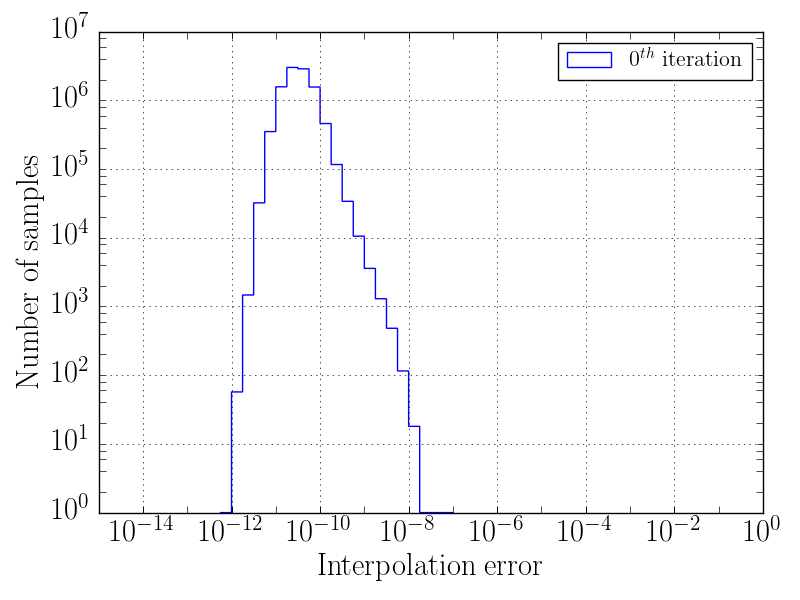}
\includegraphics[width=8cm]{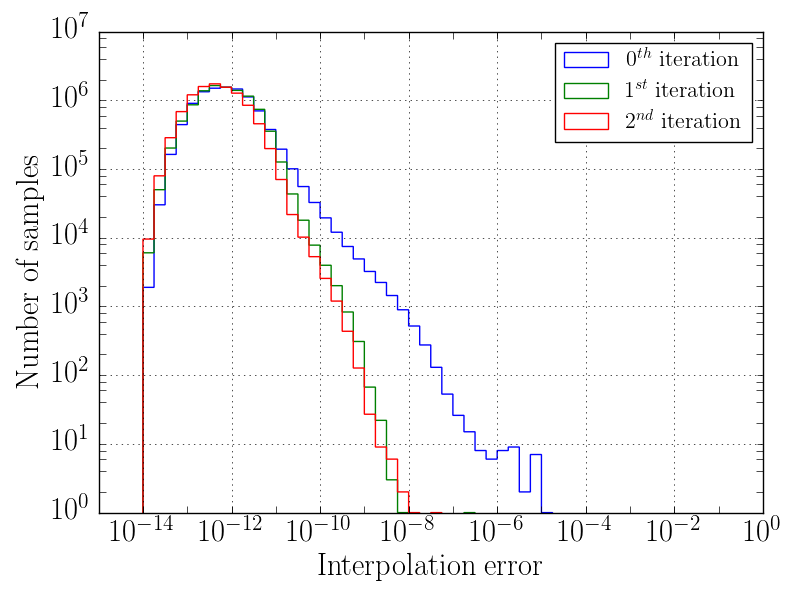}
\includegraphics[width=8cm]{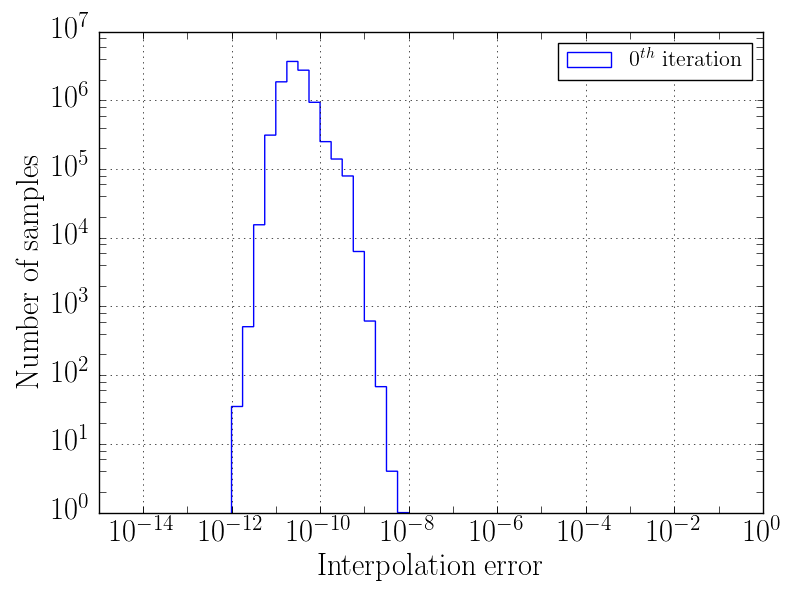}
\includegraphics[width=8cm]{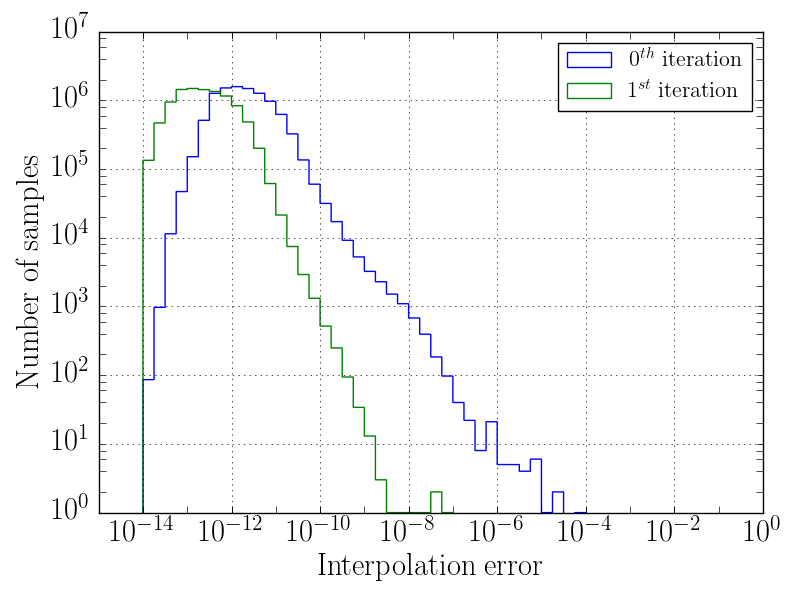}
\caption{Distributions of interpolation errors at different validation steps for 
some of the testing parameters and chirp mass bins. The left column is for the linear 
bases, the right column for the quadratic ones. First row: $\delta\hat{\varphi}_3$ for 
chirp mass bin A in Table I; second row: $\delta\hat{\beta}_2$ for bin C; third row: 
$\delta\hat{\alpha}_2$ for bin C. In some cases no ``bad points" are found, so that 
the basis does not need to be enlarged and no further validation steps are needed.}
\label{fig:interpolationerrors}
\end{figure*}

Like for the GR version of IMRPhenomPv2, the greedy projection error is set to $\epsilon = 10^{-8}$,
and the maximum interpolation error to $\beta = 10^{-6}$. Some representative distributions 
of the interpolation error at different validation steps are shown in Fig.~\ref{fig:interpolationerrors}. 
As it turns out, the addition of a
testing parameter $\delta\hat{\varphi}_i$ typically increases the sizes of the final 
bases in the different chirp mass bins by only a factor of a few, though with some exceptions;
the largest increase happens to be for $\delta\hat{\varphi}_1$ and 
$\mathcal{M} \in [3.4, 6.2]\,M_\odot$, where the linear basis size went from 
524 to 5264. 

Table II shows the speed-ups in likelihood calculations
-- defined as $\left[(f_{\rm max} - f_{\rm min}) T + 1\right]/(N_L + N_Q)$ --
that are achievable with the ROQs. The speed-up is greatest for long signals where analyses are the
most involved. These are the theoretical speed-ups; the actual speed-ups in practical parameter
estimation will vary, but tend to be the same as the theoretical ones within a factor of two or less.

\begin{table}
\begin{tabular}{ccccccccccc}
\hline
$\delta\hat{p}_i$ & \vline & A & \vline & B & \vline & C & \vline & D & \vline & E \\
\hline
\hline
$\delta\hat{\varphi}_0$ &  \vline & 4.3 & \vline & 7.6 & \vline & 28.3 & \vline & 38.7 & \vline & 47.1\\
$\delta\hat{\varphi}_1$ &  \vline &  3.0  & \vline & 4.6 & \vline & 8.6 & \vline & 11.7 & \vline & 27.1 \\
$\delta\hat{\varphi}_2$ &  \vline &  4.2  & \vline & 6.8 & \vline & 24.8 & \vline & 42.4 & \vline & 56.5 \\
$\delta\hat{\varphi}_3$ &  \vline & 4.0  & \vline & 6.1 & \vline & 20.8 & \vline & 36.9 & \vline & 55.8 \\
$\delta\hat{\varphi}_4$ &  \vline & 3.9 & \vline & 10.1 & \vline & 40.3 & \vline & 76.5 & \vline & 111.2 \\
$\delta\hat{\varphi}_{5l}$ &  \vline & 4.1 & \vline & 7.6 & \vline & 29.9 & \vline & 62.9 & \vline & 97.9 \\
$\delta\hat{\varphi}_6$ &  \vline & 3.7 & \vline & 9.8 & \vline & 39.0 & \vline & 76.0 & \vline & 114.0 \\
$\delta\hat{\varphi}_{6l}$ &  \vline & 3.8 & \vline & 10.1 & \vline & 42.1 & \vline & 78.1 & \vline & 117.1 \\
$\delta\hat{\varphi}_7$ &  \vline &  3.7 & \vline & 9.1 & \vline & 39.5 & \vline & 74.7 & \vline & 112.6 \\
$\delta\hat{\beta}_2$ &  \vline & 3.0 & \vline & 8.7 & \vline & 34.9 & \vline & 78.0 & \vline & 117.5\\
$\delta\hat{\beta}_3$ &  \vline & 3.5 & \vline & 6.8 & \vline & 28.5 & \vline & 69.8 & \vline & 111.2 \\
$\delta\hat{\alpha}_2$ &  \vline & 2.8 & \vline & 9.2 & \vline & 39.4 & \vline & 88.2 & \vline & 124.6 \\
$\delta\hat{\alpha}_3$ &  \vline & 2.9 & \vline & 10.8 & \vline & 44.8 & \vline & 87.5 & \vline & 128.3 \\
$\delta\hat{\alpha}_4$ &  \vline &  2.8 & \vline & 10.4 & \vline & 43.3 & \vline & 88.1 & \vline & 131.6 \\
\hline
\label{table:speedups}
\end{tabular}
\caption{Theoretical speed-ups of likelihood calculations due to the ROQs, for different testing parameters
and the chirp mass bins of Table I. Note how these are larger for longer signals, where they are
the most needed. Speed-ups in practical parameter estimation will vary, but tend to be the 
same as the theoretical ones within a factor of two or less.}
\end{table}

\begin{figure*}
\includegraphics[width=8cm]{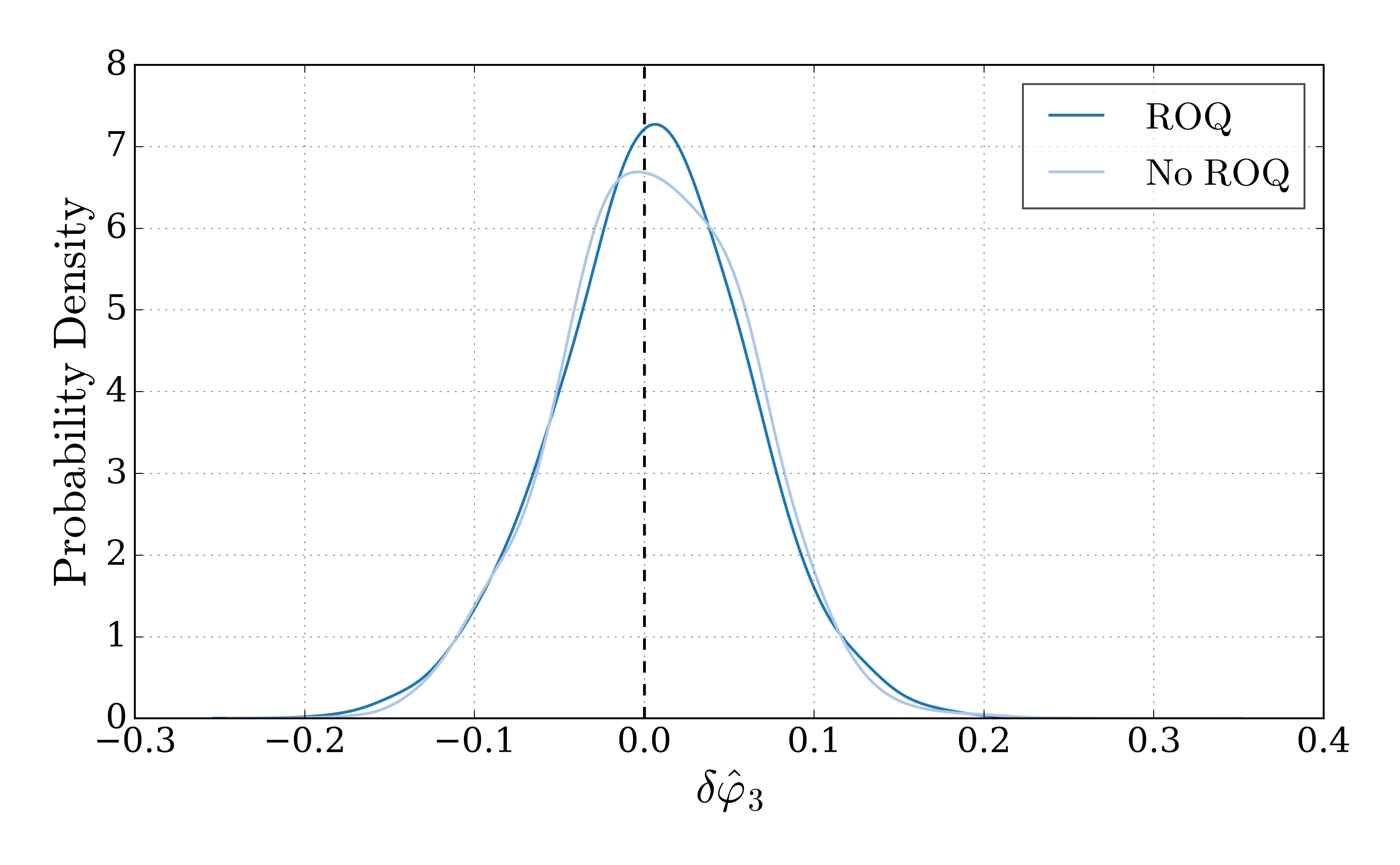}
\includegraphics[width=8cm]{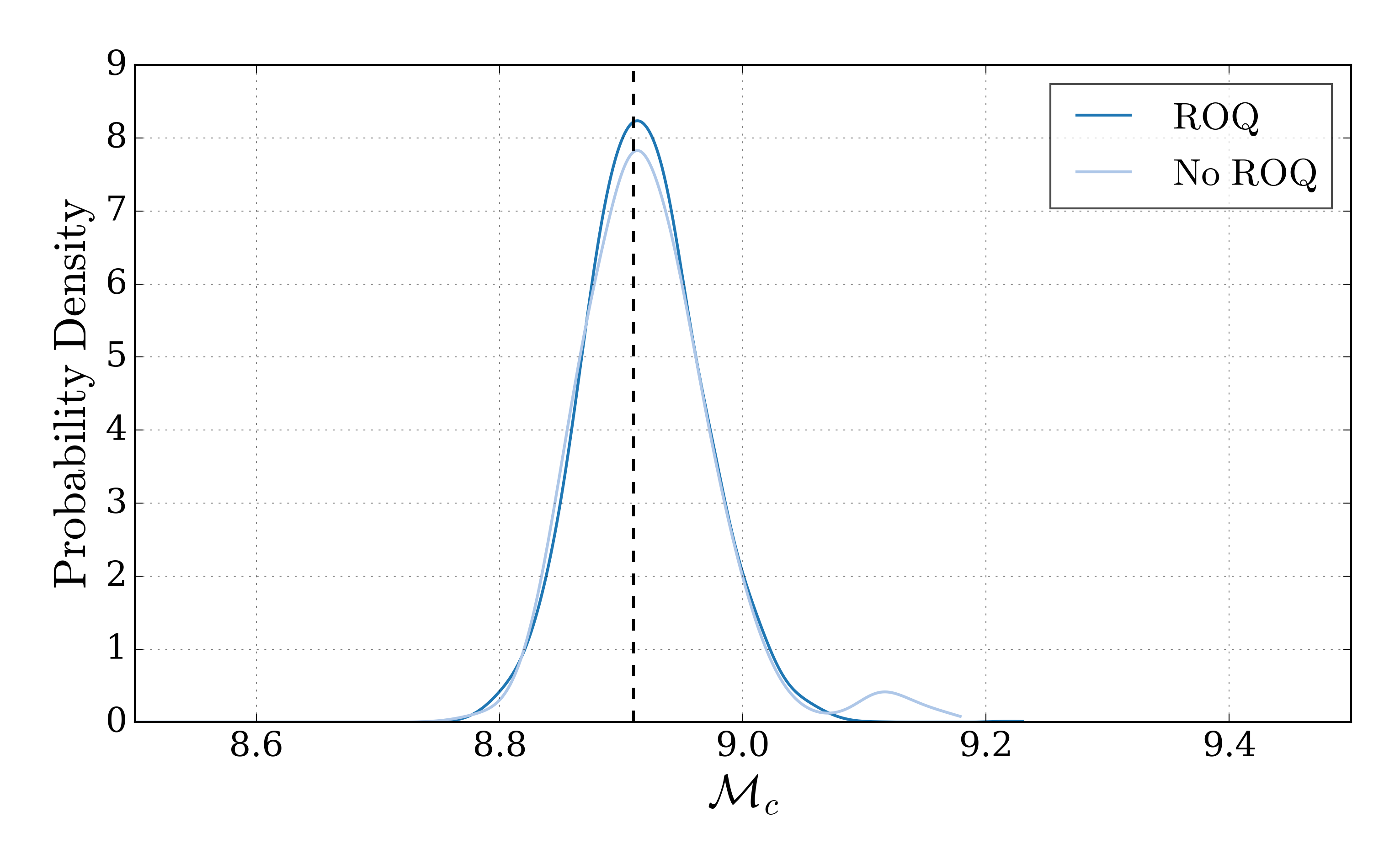}
\includegraphics[width=8cm]{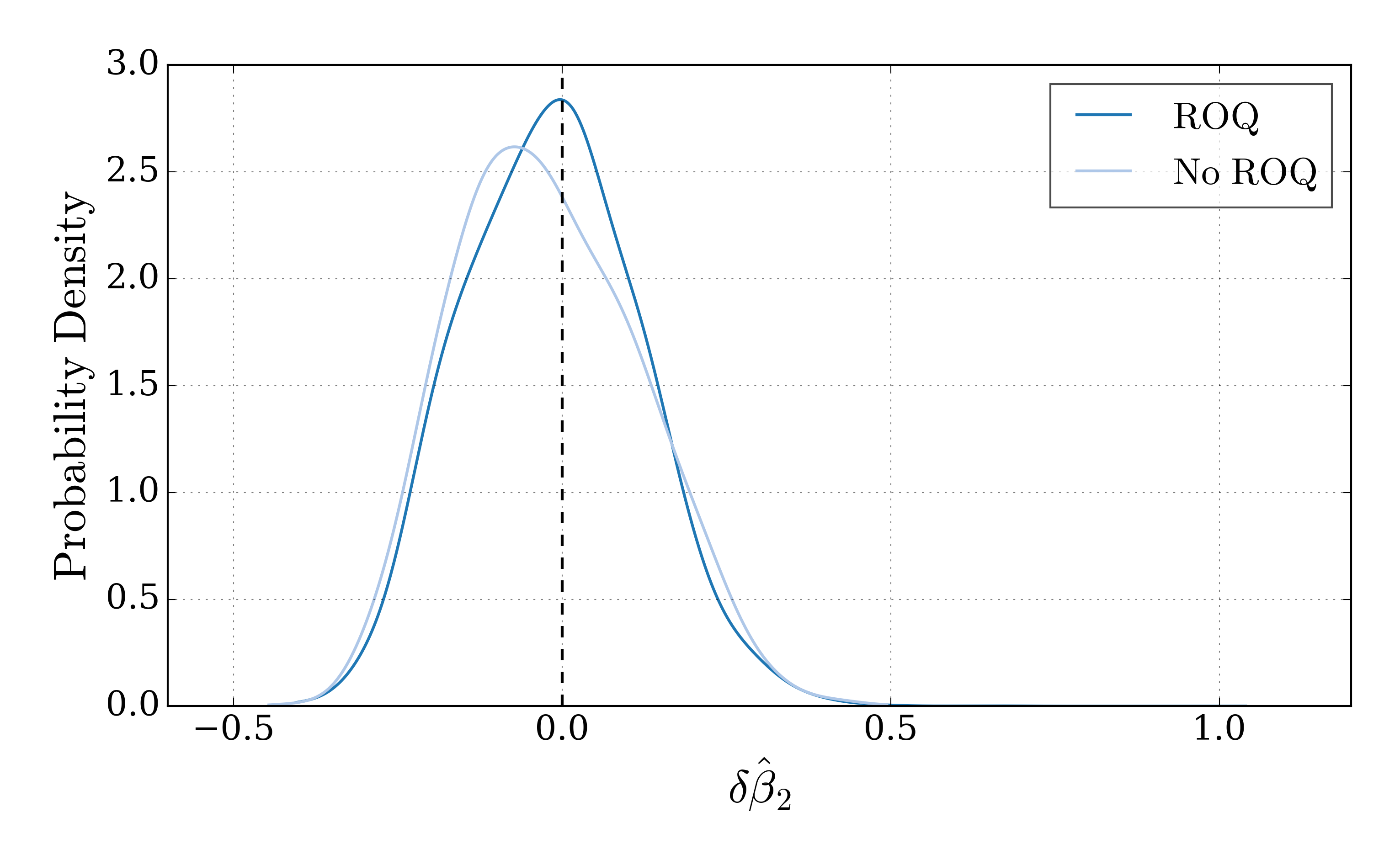}
\includegraphics[width=8cm]{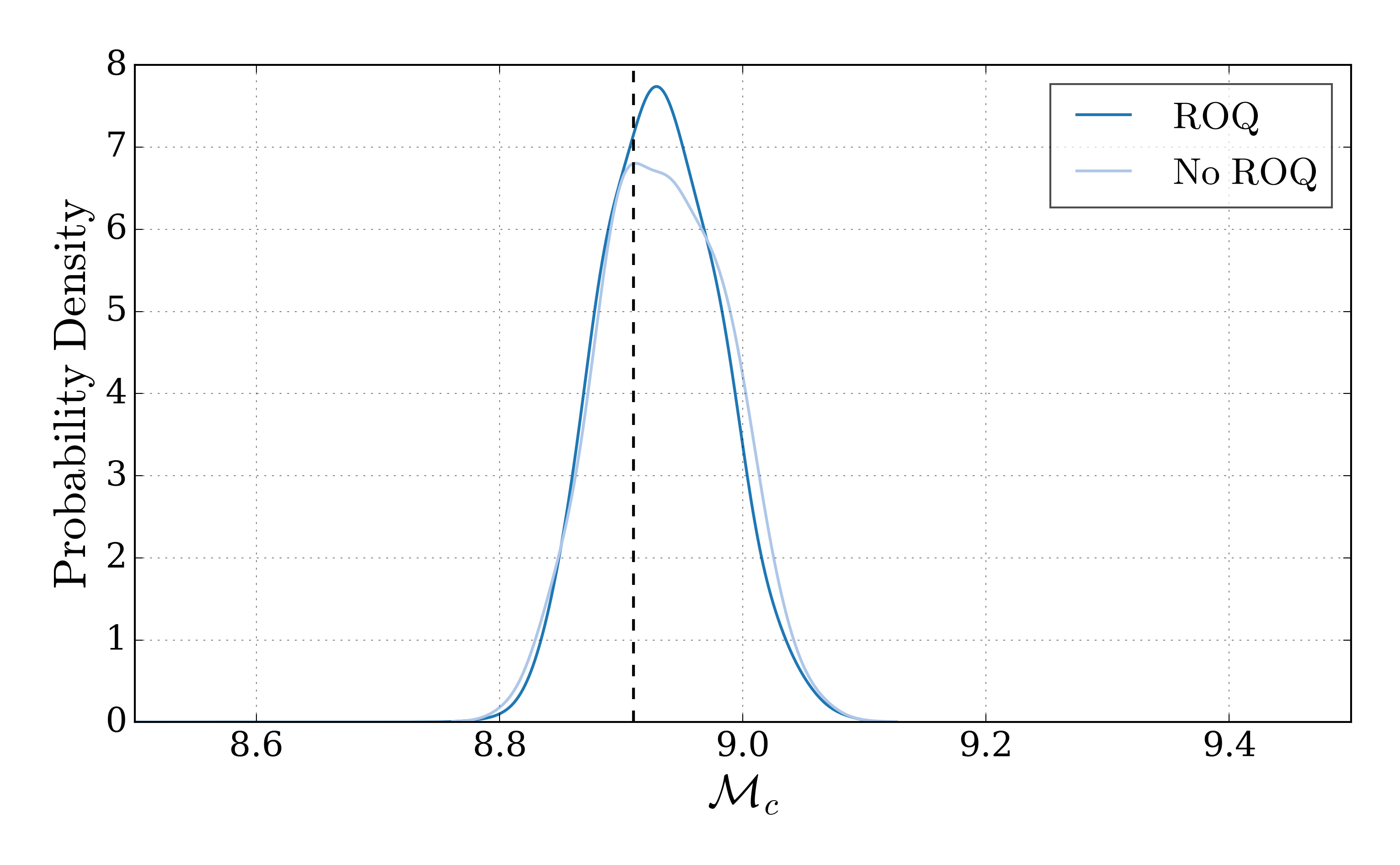}
\includegraphics[width=8cm]{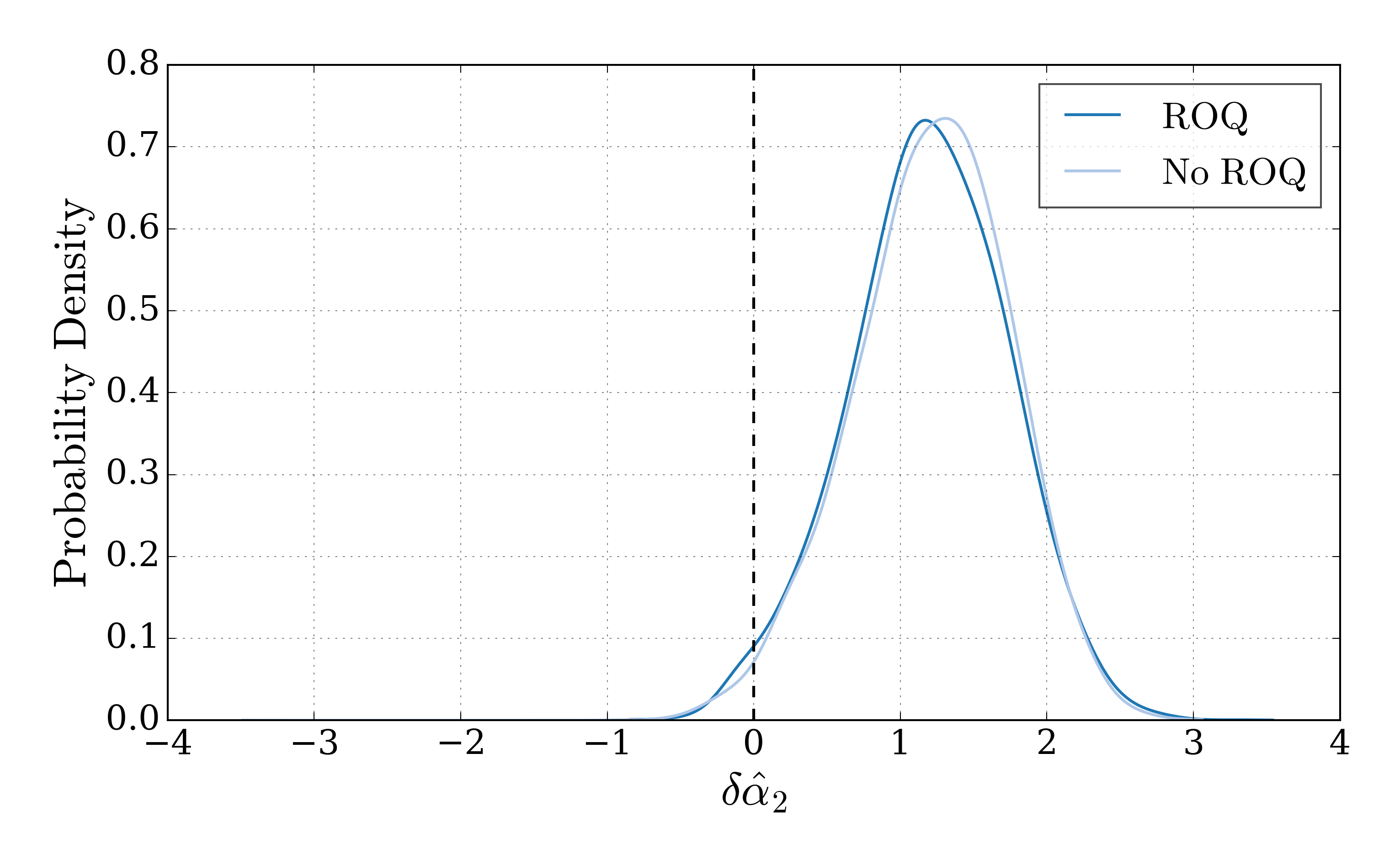}
\includegraphics[width=8cm]{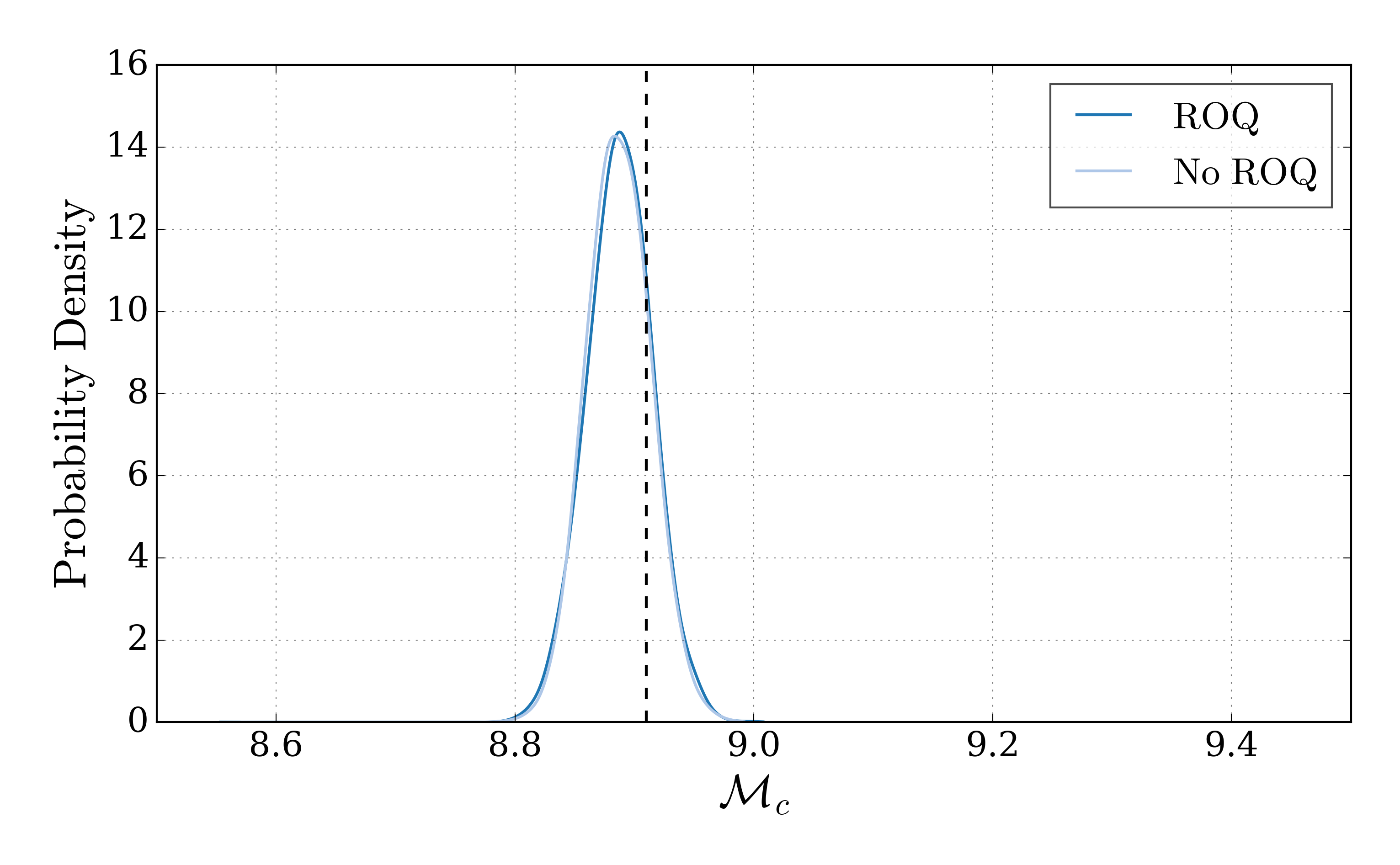}
\caption{A comparison of parameter estimation results on a simulated signal 
with parameters $\mathcal{M}_c = 8.9\,M_\odot$, $q = 1.99$, $D_{\rm L} = 200$ Mpc, 
and $\delta\hat{p}_i = 0$, in synthetic stationary, Gaussian noise, 
analyzed with and without the ROQ. Results are shown for the 
cases $\delta\hat{\varphi}_3$ (top row), $\delta\hat{\beta}_2$ (middle row), and
$\delta\hat{\alpha}_2$ (bottom row). In each case we show the posterior density function
for the testing parameter itself (left column) and for chirp mass (right column). 
The values of the parameters in the signal are indicated by the vertical dashed lines.
Results with and without ROQ agree to within sampling uncertainties \cite{Veitch:2014wba}.}
\label{fig:PE_ROQ_noROQ}
\end{figure*}

Finally, the ROQs were interfaced with the abovementioned LALInference framework. 
Fig.~\ref{fig:PE_ROQ_noROQ} compares some parameter estimation results obtained
with and without the ROQ on the same simulated signal. We see that the results 
are consistent, with posterior density functions not differing by more than 
what is expected given uncertainties in the sampling process \cite{Veitch:2014wba}.
The robustness of the infrastructure will be tested in more detail in the next 
section.

\section{Robustness of the tests}

We now perform some checks of the correctness of the data analysis pipeline, 
and its robustness against waveform systematics and instrumental noise. We do this 
in two ways. One is to construct so-called $p$-$p$ plots, which quantify the statistical 
inconsistencies of the posterior density distributions. Another consists of analyzing a numerical 
waveform injected in many different stretches of real detector noise, as a check that the 
pipeline behaves as it should under the combined effects of the injected waveform
being different from the template waveform model, and the presence of instrumental
glitches in the detector output. 

\subsection{Reliable measurement of testing parameters}
\label{subsec:pp_plots}

\begin{figure*}
\includegraphics[width=5cm]{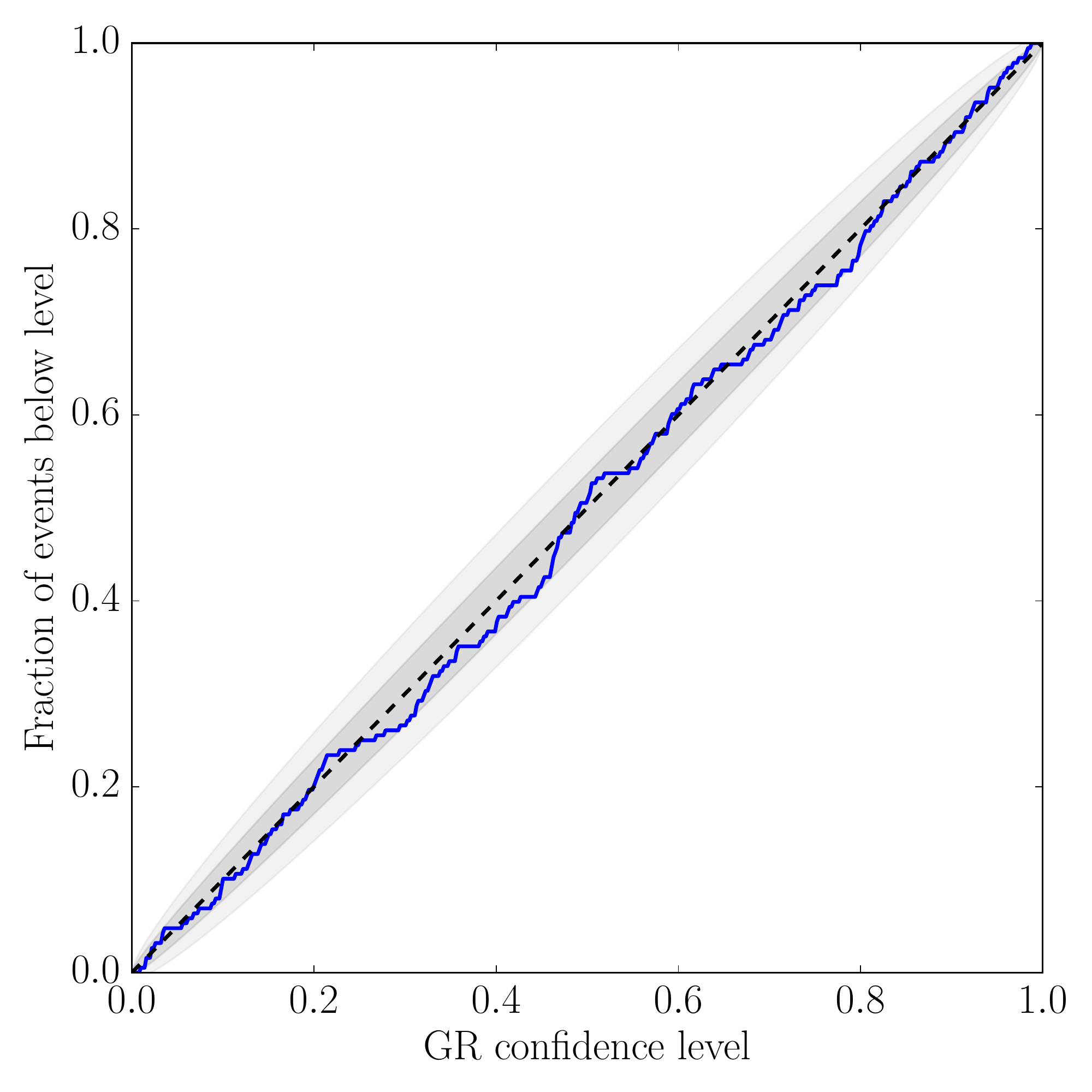}
\includegraphics[width=5cm]{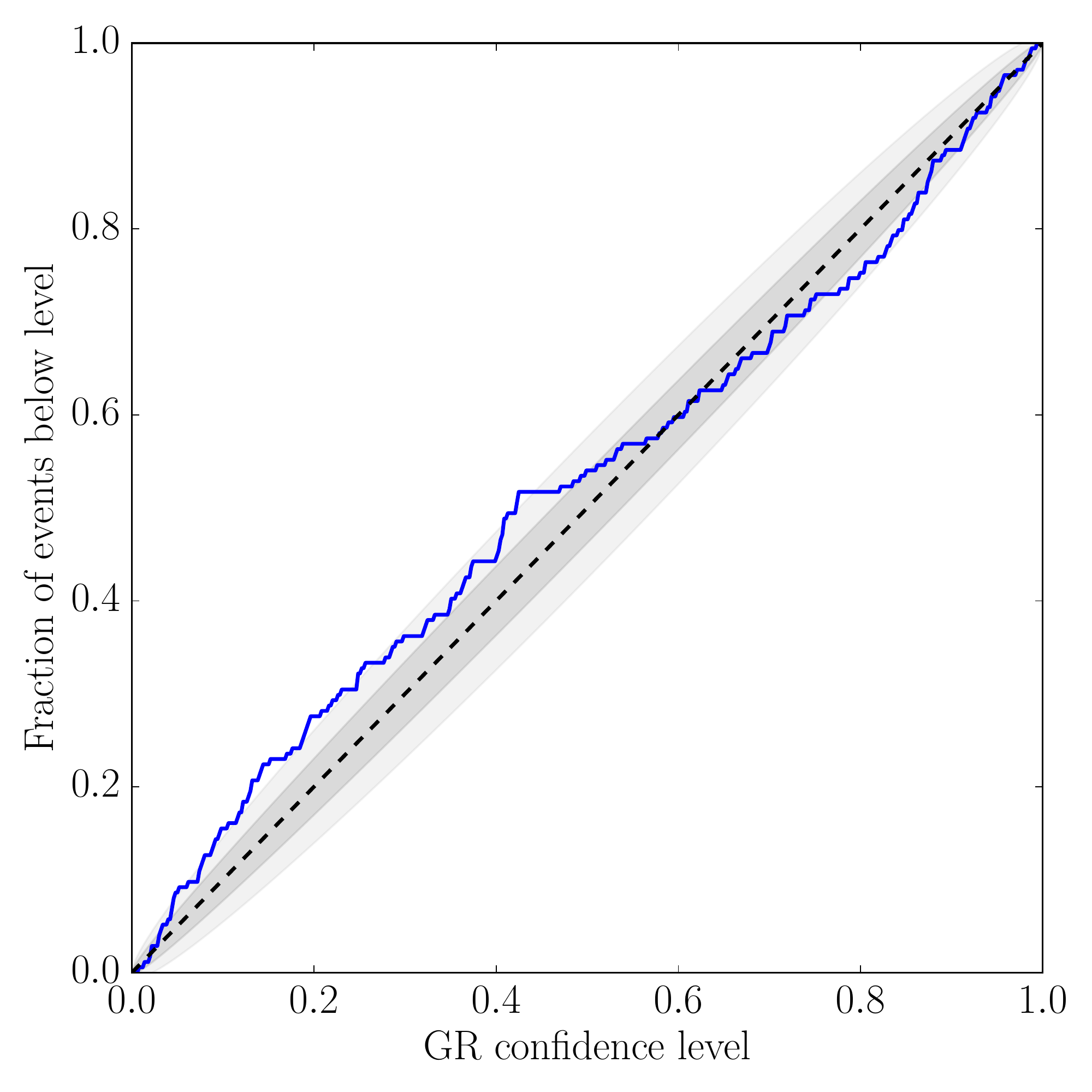}
\includegraphics[width=5cm]{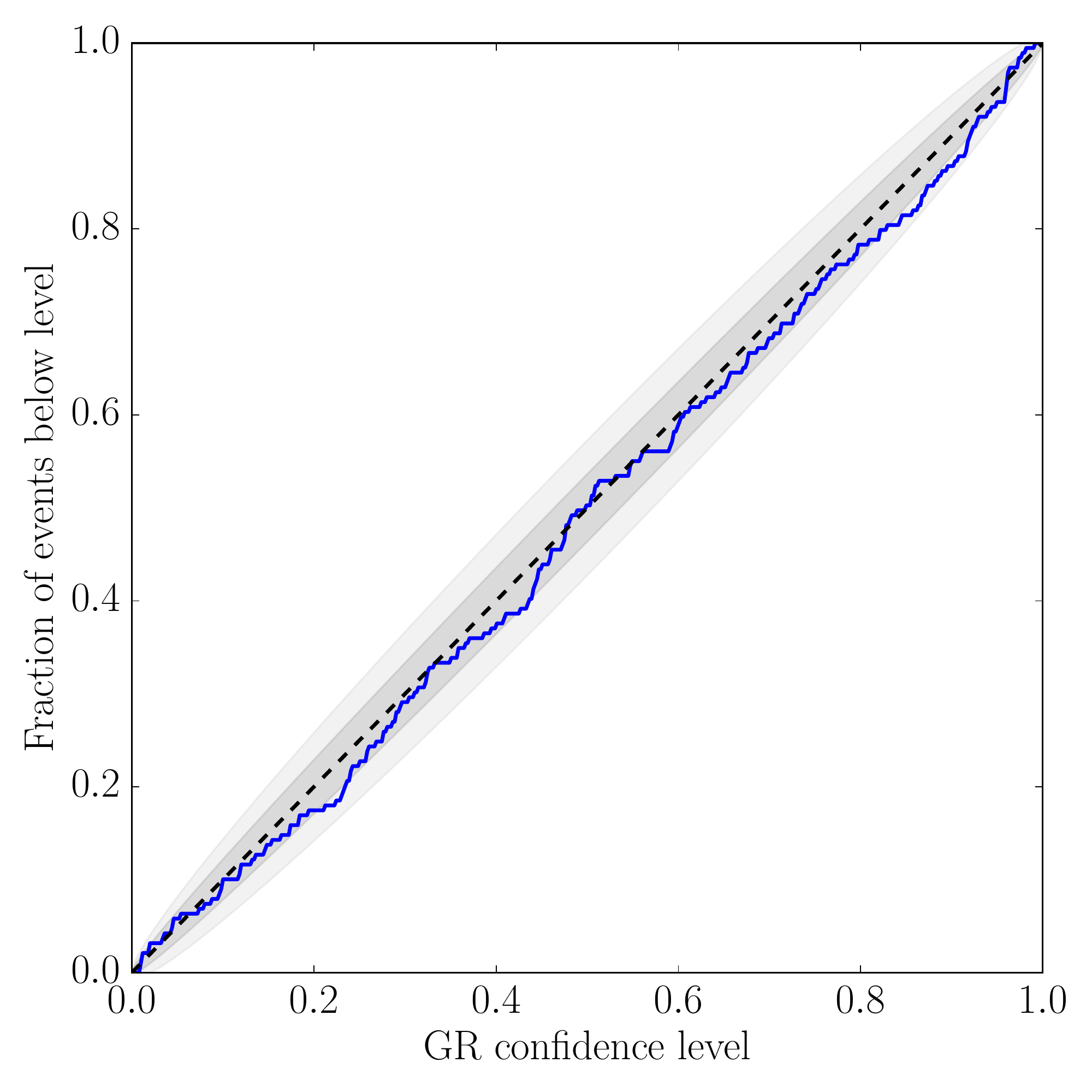}
\caption{Fraction of simulated signals in stationary, Gaussian noise for which the value of 
zero for $\delta\hat{p}_i$ is within a given confidence level. Shown are $p$-$p$ plots
for $\delta\hat{\varphi}_3$ (left), $\delta\hat{\beta}_2$ (middle), and $\delta\hat{\alpha}_2$
(right). The dark and light gray bands indicate the 1-$\sigma$ and 2-$\sigma$ 
departures from the diagonal that can be expected on theoretical grounds. The results 
are consistent with a general absence of bias in the measurements.}
\label{fig:PPplots}
\end{figure*}

A requirement for a parameter estimation algorithm is that it is capable of measuring parameters
in a statistically reliable way. Detector noise can cause offsets in posterior density 
functions, but given a large number of signals it should be the case that the correct 
parameter value is recovered with a confidence $p$ in a fraction $p$ of the cases. 
Specifically, assuming GR is correct, for any of the parameterized tests it should be the case that
the value $\delta\hat{p}_i = 0$ lies in a confidence interval of width $p$ for a fraction
$p$ of the measurements. We check this by adding 100 simulated GR signals (\emph{injections}) to synthetic, 
stationary, Gaussian noise, with the predicted power spectral density at design sensitivity 
for the two Advanced LIGO detectors \cite{AdvLIGOPSD}. The signals have randomly chosen sky positions
and orientations and are placed uniformly in co-moving volume with $D_{\rm L} \in [250, 750]$ Mpc, 
with component masses $m_1, m_2 \in [6,40]\,M_\odot$, and arbitrarily oriented spins
with magnitudes $|\mathbf{S}_1|, |\mathbf{S}_2| \in [0, 0.9]$. Injections are analyzed with 
the ROQs whose chirp mass bins they fall into; in reality one would look at the 
chirp mass measured with GR templates. $p$-$p$ plots for a few of the testing parameters
are shown in Fig.~\ref{fig:PPplots}. As an indicator of consistency of the results 
with absence of bias in the measurements, one can calculate the Kolmogorov-Smirnov (K-S) 
statistic, which is defined as the maximum (in absolute value) of the difference
between distributions; in our case the latter are simply the $p$-$p$ distribution on the
one hand, and the diagonal on the other. We find K-S values of 0.04, 0.09, and 0.04 for
$\delta\hat{\varphi}_3$, $\delta\hat{\beta}_2$, and $\delta\hat{\alpha}_2$ respectively. We 
conclude that the analyses work as expected.

\subsection{Numerical relativity injections in real detector noise}

Finally we investigate the response of the parameterized tests to a numerical relativity waveform 
injected in detector noise that contains instrumental glitches. In particular, we use 
real data from the S6 data set, but ``recolored" to the early advanced Advanced LIGO 
noise curve from \cite{Aasi:2013wya}; this procedure changes the average power spectral density 
but retains (and in fact enhances) any instrumental non-stationarities that were present 
in the original data. Since instrumental glitches will have a larger effect for short-duration 
signals, we focus on GW150914. We consider a numerical relativity waveform 
from the SXS catalog, whose mass ratio and spins are close to the measured means for 
GR150914; specifically, we pick SXS:BBH:0307 \cite{sxs}. The intrinsic parameters
were $(m_1, m_2) = (40.83, 33.26)\,M_\odot$, and $\mathbf{S}_1 = (0.092,0.038,0.326)$, 
$\mathbf{S}_2 = (0.215, 0.301, -0.558)$ at $f_{\rm ref} = 20$ Hz. This same waveform is then 
injected in 21 different stretches of noise \cite{Galley:2016mvy,Schmidt:2017btt}, and the 
parameterized tests are performed.
In choosing these stretches, care was taken to pick ones that did not exhibit egregiously
large glitches (which can be done by visual inspection of time-frequency spectrograms), since
the presence of a sufficiently sizeable departure from Gaussianity of the noise 
may preclude an event being detected in the first place. The strategy is 
similar to what was followed in \cite{Abbott:2016wiq} (see their Sec.~III.E), where the effect of 
possible non-stationarities on parameter estimation -- in the GR case -- was also assessed by injecting a 
particular numerical relativity waveform in different stretches of real detector noise.

As a diagnostic we define the ``GR quantile" as the cumulative probability of
a given $\delta\hat{p}_i$ being non-positive:
\be
Q_i \equiv \int_{-\infty}^0 p(\delta\hat{p}_i|H_i, d, I)\,d\delta\hat{p}_i.
\ee
If the GR quantile is close to zero then the posterior $p(\delta\hat{p}_i|H_i, d, I)$
exhibits a significant offset towards positive $\delta\hat{p}_i$; if it is close to 
one then there is a large offset towards negative values. Given many measurements on
the same signal in different noise realizations, we expect the $Q_i$ to be distributed
uniformly on the interval $[0,1]$. 

\begin{figure*}
\includegraphics[width=18cm]{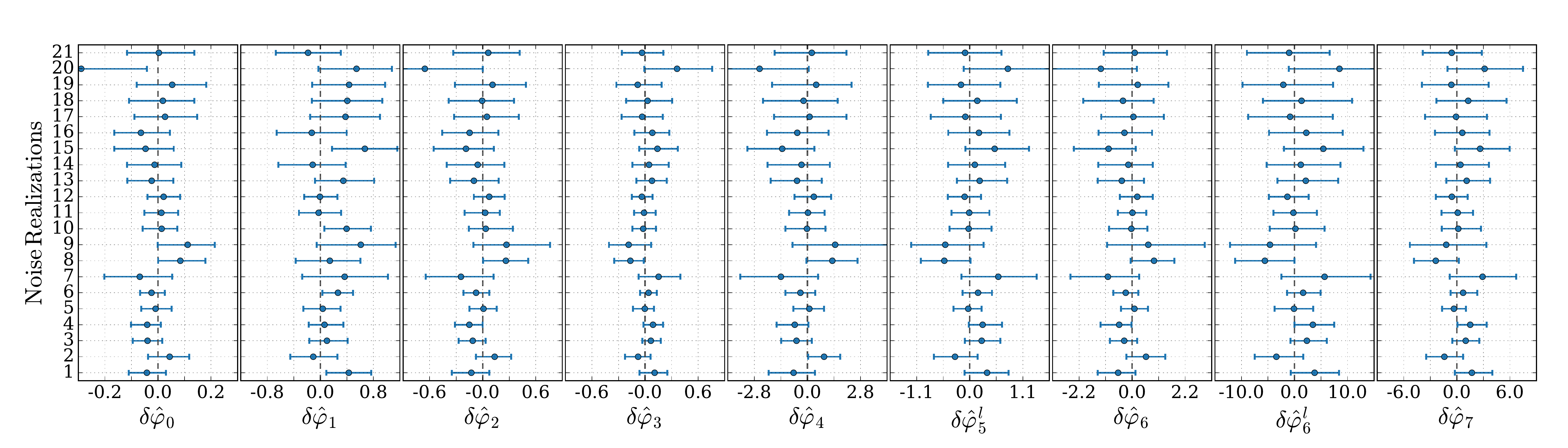}
\caption{90\% credible intervals for the PN testing parameters obtained by
performing the parameterized tests on a numerical relativity injection in 21 different
stretches of realistic detector data. Note how offsets tend to alternate from one
PN testing parameter to the next; this is due to partial correlation between them, and the
alternating signs of the PN parameters themselves.}
\label{fig:confidenceintervals}
\end{figure*}

\begin{figure}
\includegraphics[width=\columnwidth]{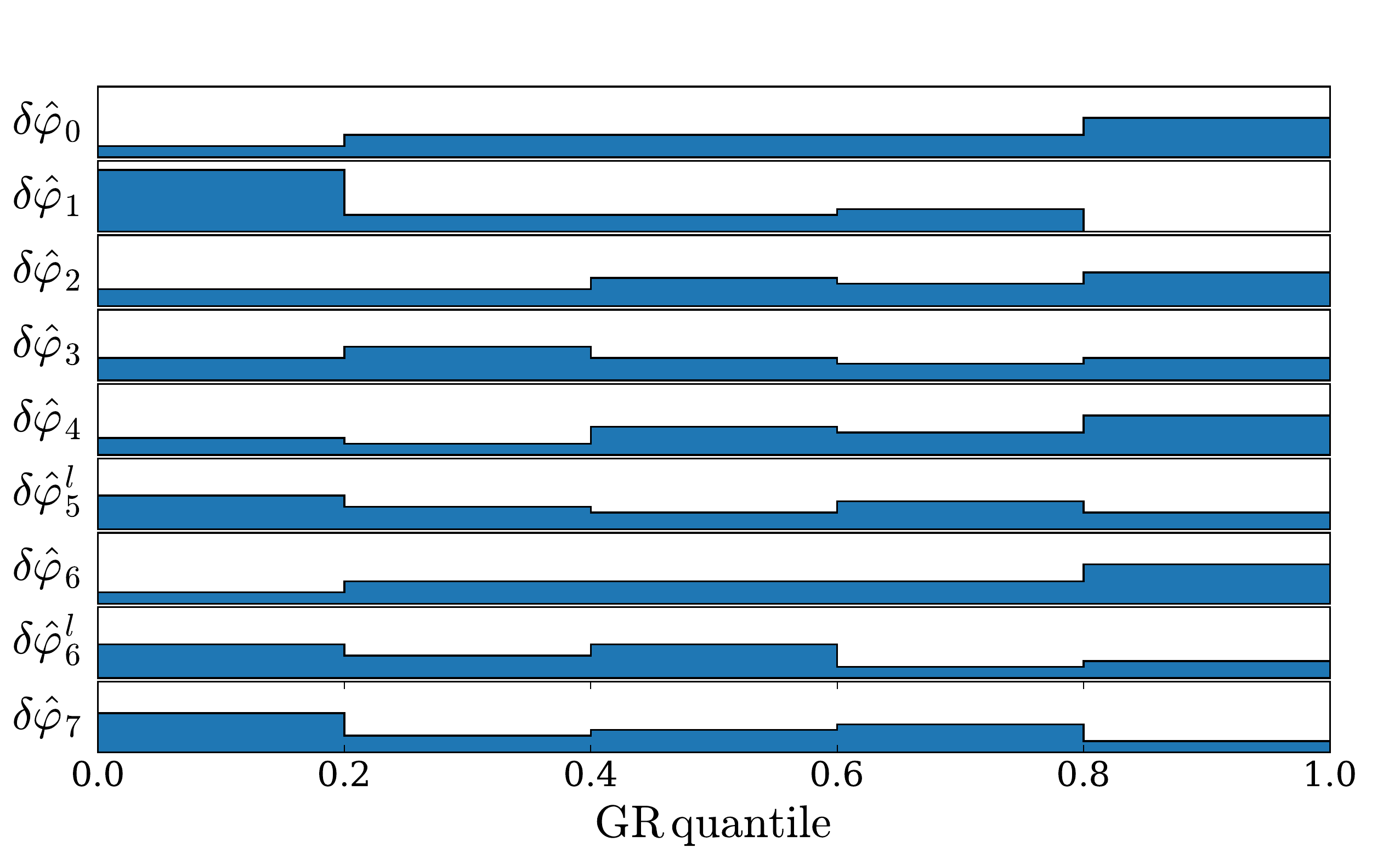}
\caption{Histograms of GR quantiles for the PN testing parameters corresponding to
the same simulations as for Fig.~\ref{fig:confidenceintervals}. Though based on
analyses of only 21 stretches of data, the results are consistent with the quantiles
being uniformly distributed on the interval $[0,1]$.}
\label{fig:quantiles}
\end{figure}

In Fig.~\ref{fig:confidenceintervals} we first of all show the 90\% credible
intervals for the PN testing parameters 
$\{\delta\hat{\varphi}_0, \ldots, \delta\hat{\varphi}_7\}$ and 
$\{\delta\hat{\varphi}_{5l}, \delta\hat{\varphi}_{6l}\}$ for the 21 stretches
of data. We note how the deviations in the PN parameters tend to alternate in sign, 
due to the fact that there is some correlation between them, and that the $\varphi_i$
themselves have alternating signs. Next, in Fig.~\ref{fig:quantiles} we show the
distribution of the $Q_i$, which despite the small sample size is indeed suggestive of
uniformity on $[0,1]$.

Needless to say, a full investigation for systems like GW150914 would require performing 
the parameterized tests for a much larger sample of data stretches than the 21 used here, 
and it would be of interest to repeat the study for other choices of masses and spins; due to
computational restrictions this was not practicable. Nevertheless, the outcome is 
indicative of the expected behavior.

\section{Measurement sensitivities}

Next we want to assess the power of our parameterized tests in constraining 
GR violations, and their sensitivity to selected GR violations, by adding simulated
signals to stationary, Gaussian detector noise with the
power spectral density of the Advanced LIGO detectors at design sensitivity \cite{AdvLIGOPSD}, 
and performing parameter estimation as in the previous section. 

As far as GR violations are concerned, ideally one would like to do this using specific 
alternative theories of gravity. 
However, in most cases the effects of particular theories have only been calculated
for the inspiral, and then only to leading PN order 
\cite{YunesPretorius:2009,Cornish:2011ys,Yunes:2013dva,Yunes:2016jcc}; to our knowledge, full 
inspiral-merger-ringdown waveform models with reasonable inclusion of 
all relevant physical effects so far only exist for GR itself. Hence we confine
ourselves to injections that have a deviation $\delta\hat{p}_i$ in a particular coefficient
$p_i$, or in several of the $p_i$ at the same time, starting from some PN order. However,
in the template waveforms used for the measurements, we still only vary a single 
one of the $\delta\hat{p}_i$ at a time. As we shall see, if the injections have 
deviations in multiple coefficients, then single-parameter tests will still pick this up. 
In fact, even parameters that are not associated with the deviations in the signal must 
show deviations. Such effects had already been observed in 
\cite{Li:2011cg,Li:2011vx,Sampson:2013lpa}, and should 
not come as a surprise: template waveform models will use whatever additional freedom they have
to accommodate anomalies in the signals. At the same time, only varying one testing parameter
leads to a higher measurement accuracy than for multiple parameters being varied at the same time.
A drawback is that posterior densities for testing parameters 
can not be straightforwardly mapped to statements about whatever additional charges, coupling constants, 
or energy scales may be present in some particular alternative theories. 
For this to be possible, accurate and complete inspiral-merger-ringdown waveforms for
alternative theories would be required, but these are not currently available. 
However, the purpose of the parameterized tests is not to place bounds on parameters
characterizing other theories, but rather to test the theory of general relativity 
itself, with as high an accuracy as possible.

\subsection{Bounding GR violations}

\begin{figure*}
\includegraphics[width=5.5cm]{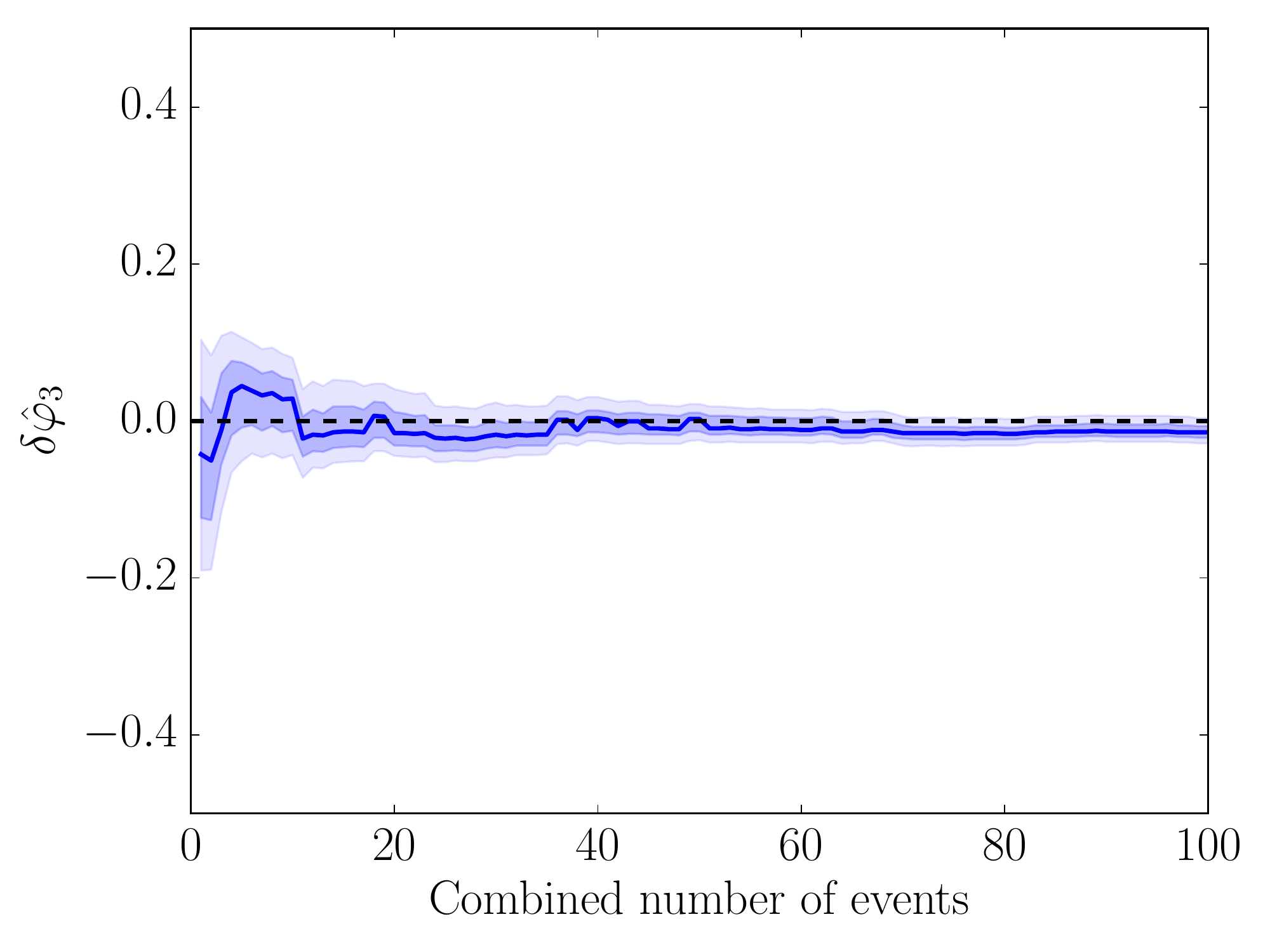}
\includegraphics[width=5.5cm]{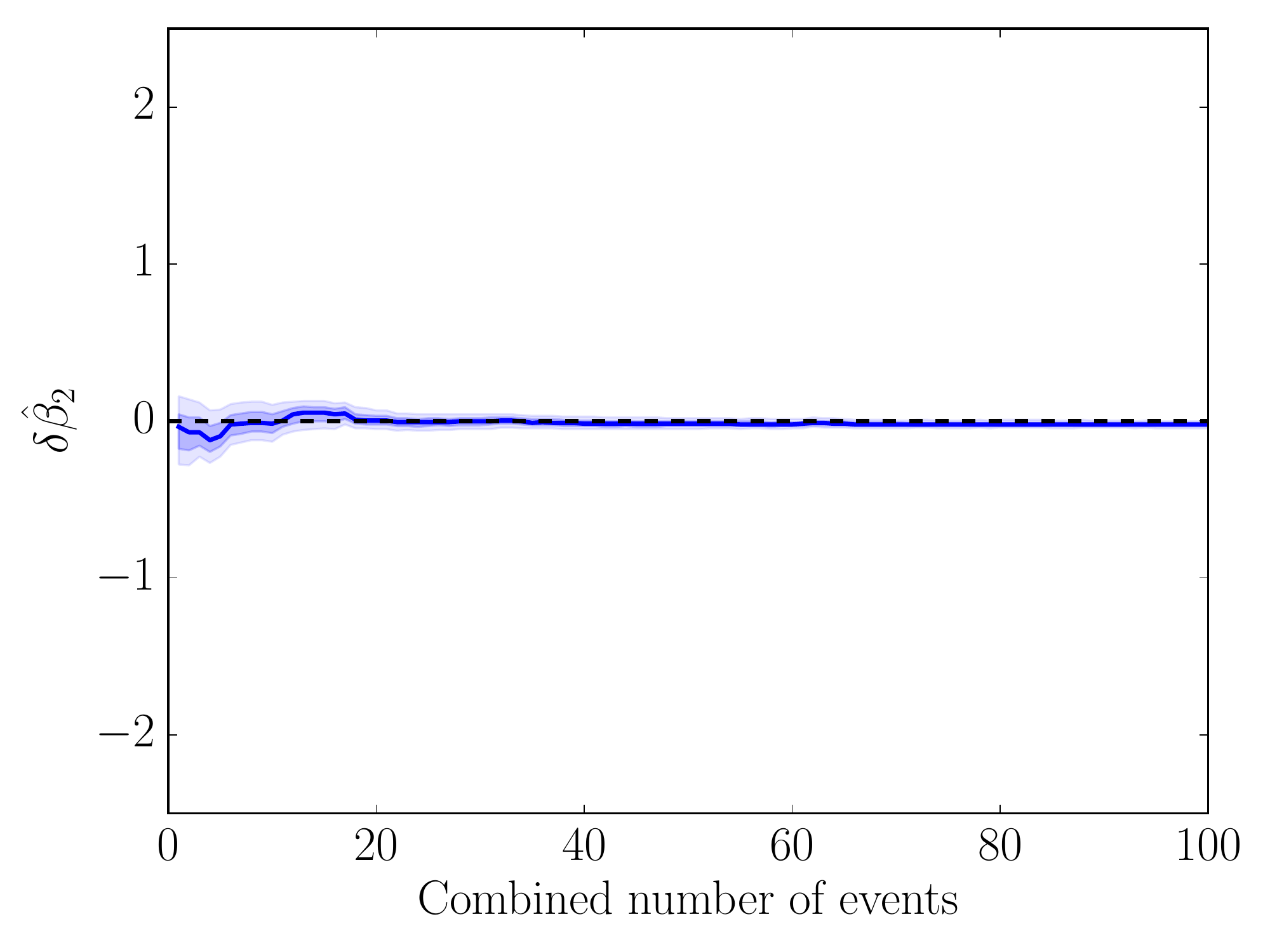}
\includegraphics[width=5.5cm]{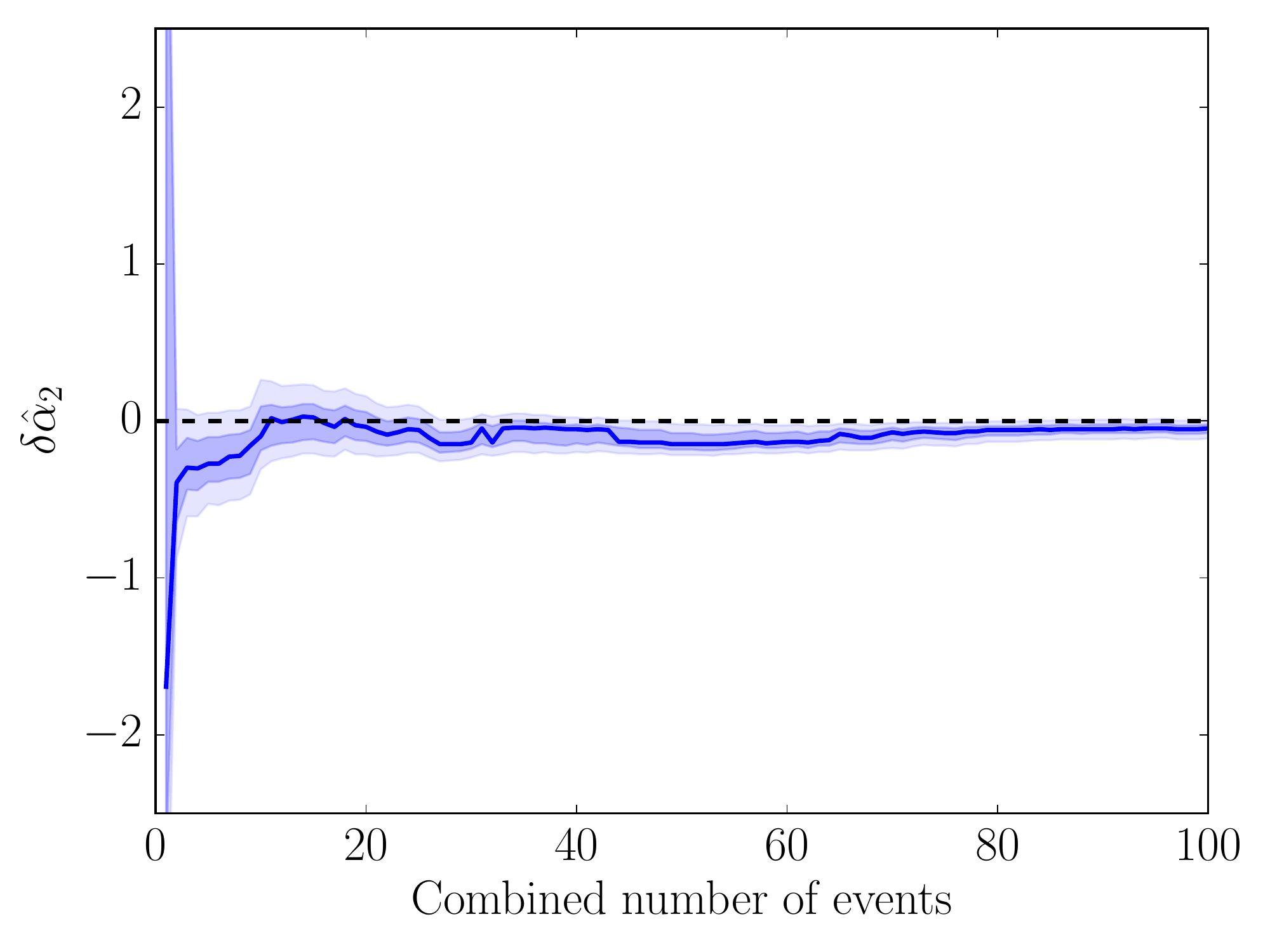}
\caption{Sharper constraints on deviation from GR can be obtained by combining
posterior density functions for the $\delta\hat{p}_i$ from all available detections. 
This is illustrated for $\delta\hat{\varphi}_3$ (left), $\delta\hat{\beta}_2$ (middle), and
$\delta\hat{\alpha}_2$ (right). The black curve shows the median of the joint 
distribution, and the darker and lighter shadings the 68\% and 95\% confidence 
intervals, respectively.}
\label{fig:combinedposteriors}
\end{figure*}

First we illustrate the ability of the parameterized tests in putting bounds on GR
violations, which will get increasingly sharper as information from multiple events
is combined. The posterior density functions for each of the $\delta\hat{p}_i$ 
obtained from the simulated signals in subsection \ref{subsec:pp_plots} lead to combined posterior
densities according to the prescription of Eq.~(\ref{combinedposteriors}). As shown in 
Fig.~\ref{fig:combinedposteriors}, after a few tens of detections these will be
sharply peaked near the value of zero. In these examples, after 50 (100) detections, 
the 1-$\sigma$ accuracies on $\delta\hat{\varphi}_3$, $\delta\hat{\beta}_2$, and
$\delta\hat{\alpha}_2$ are, respectively, 0.013 (0.008), 0.020 (0.013), and 0.054 (0.032).

\subsection{Simulated signals with deviations in particular coefficients}

We now consider injections that have a deviation in a particular
coefficient $p_i$. The GR parameters are picked to be the 
means of the posterior density distributions for GW150914 \cite{Abbott:2017xlt}. We focus on 
this type of source so 
as to have some amount of sensitivity to each of the inspiral, intermediate,
and merger-ringdown regimes. The injections are done in stationary, Gaussian noise 
with the predicted power
spectral density at design sensitivity for the two Advanced LIGO detectors \cite{AdvLIGOPSD}. 
For the deviations, we consider in turn two 
representative parameters from each of the inspiral, intermediate, and merger-ringdown 
regimes, and give the corresponding $\delta\hat{p}_i$ a magnitude that roughly corresponds
to 5 times the standard deviation observed for GW150914, with both positive and 
negative signs. In particular, $\delta\hat{\varphi}_3 = \pm 0.4$, 
$\delta\hat{\varphi}_4 = \pm 3.3$, $\delta\hat{\beta}_2 = \pm 0.7$, 
$\delta\hat{\beta}_3 = \pm 0.8$, $\delta\hat{\alpha}_2 = \pm 1.3$, and
$\delta\alpha_4 = \pm 1.6$.

\begin{figure*}
\includegraphics[width=\textwidth]{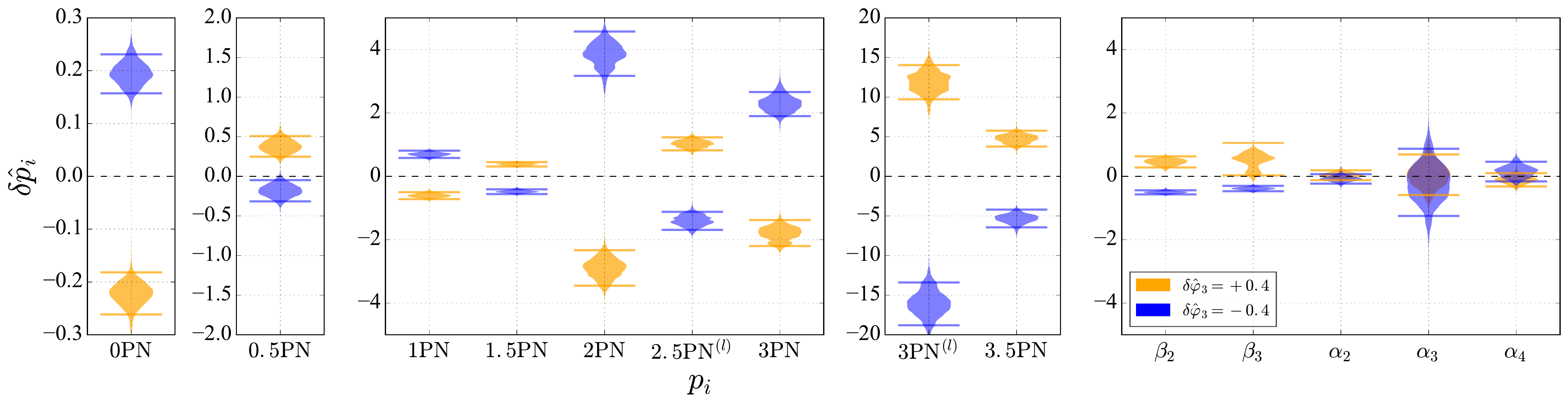}
\includegraphics[width=\textwidth]{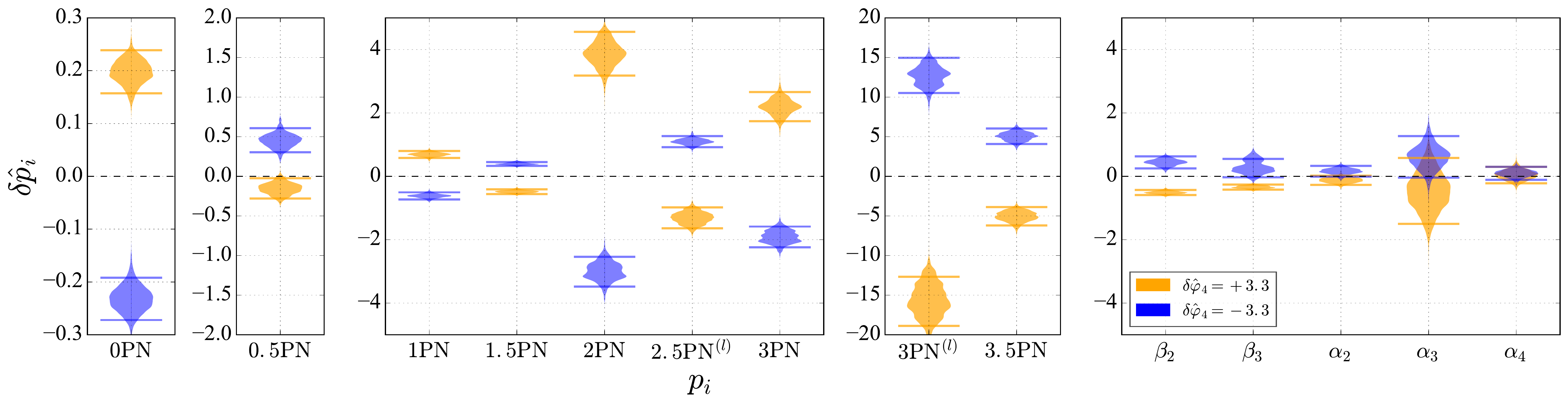}
\caption{Top: Posterior densities for testing parameters for an injection
with $\delta\hat{\varphi}_3 = +0.4$ (orange) and $\delta\hat{\varphi}_3 = -0.4$ (blue).
Bottom: posteriors for an injection with 
$\delta\hat{\varphi}_4 = +3.3$ (orange) and $\delta\hat{\varphi}_4 = -3.3$ (blue). 
Note how \emph{all} the PN testing parameters indicate a deviation from GR, not just the
ones that deviate from zero in the signal.}
\label{fig:dchis}
\end{figure*}

Fig.~\ref{fig:dchis} shows posterior densities for the cases where
the injection has either non-zero $\delta\hat{\varphi}_3$ or non-zero $\delta\hat{\varphi}_4$,
and in the measurements all of the $\delta{p}_i$ are allowed to vary in turn. A few
things can be noted:
\begin{itemize}
\item In each case, the posterior density for the testing parameter where the deviation in the
signal resides has no support at the GR value of zero, but the support does contain
the injected value. 
\item The posterior densities of \emph{all} of the other PN testing parameters, with
the exception of $\delta\hat{\varphi}_1$, show strong offsets away from zero.
\item On the other hand, the intermediate-regime and merger-ringdown testing parameters
show much less of a response to a deviation in a PN parameter.
\item The deviations in the PN parameters tend to alternate in sign. This reflects
the fact that there is some amount of correlation between these parameters, and 
that the $\varphi_i$ themselves have alternating signs.
\end{itemize}

\begin{figure*}
\includegraphics[width=\textwidth]{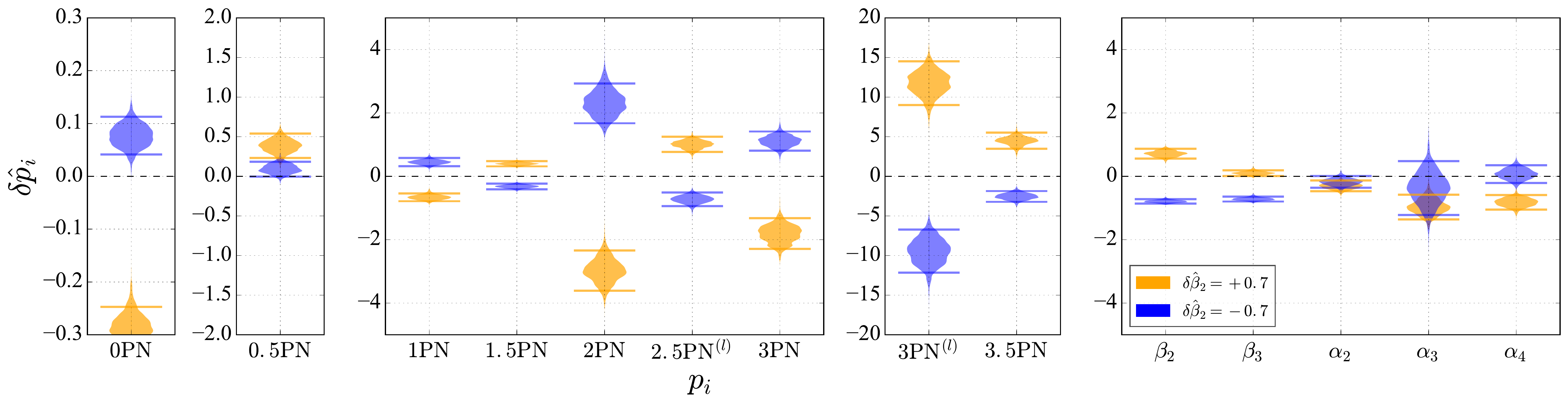}
\includegraphics[width=\textwidth]{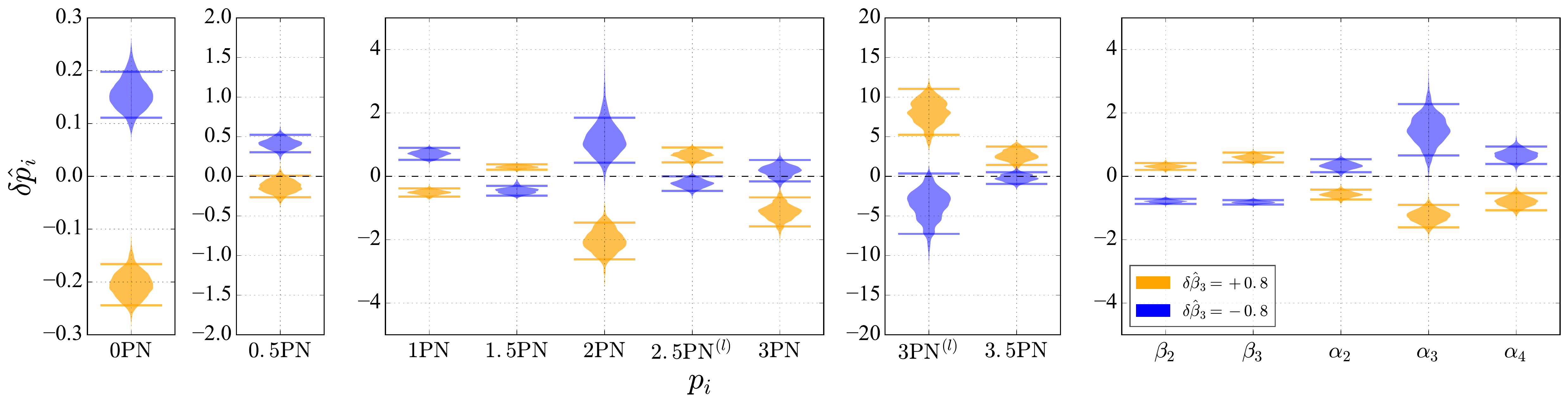}
\caption{Top: Posterior densities for testing parameters for an injection
with $\delta\hat{\beta}_2 = +0.7$ (orange) and $\delta\hat{\beta}_2 = -0.7$ (blue).
Bottom: posteriors for an injection with 
$\delta\hat{\beta}_3 = +0.8$ (orange) and $\delta\hat{\beta}_3 = -0.8$ (blue). In each
case the GR violation is also picked up by the \emph{other} $\delta\hat{\beta}_i$.}
\label{fig:dbetas}
\end{figure*}

\begin{figure*}
\includegraphics[width=\textwidth]{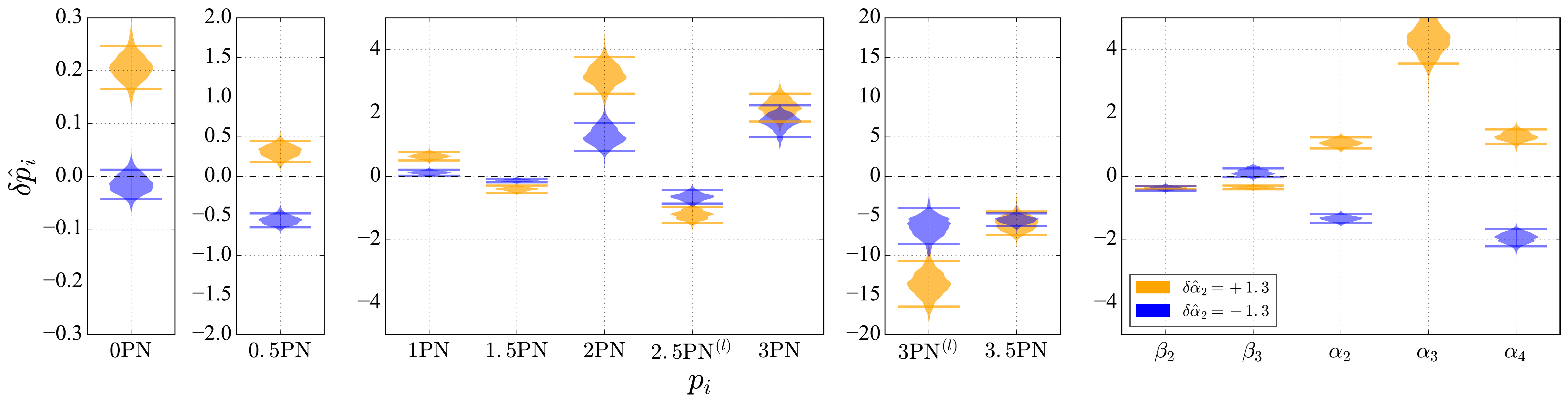}
\includegraphics[width=\textwidth]{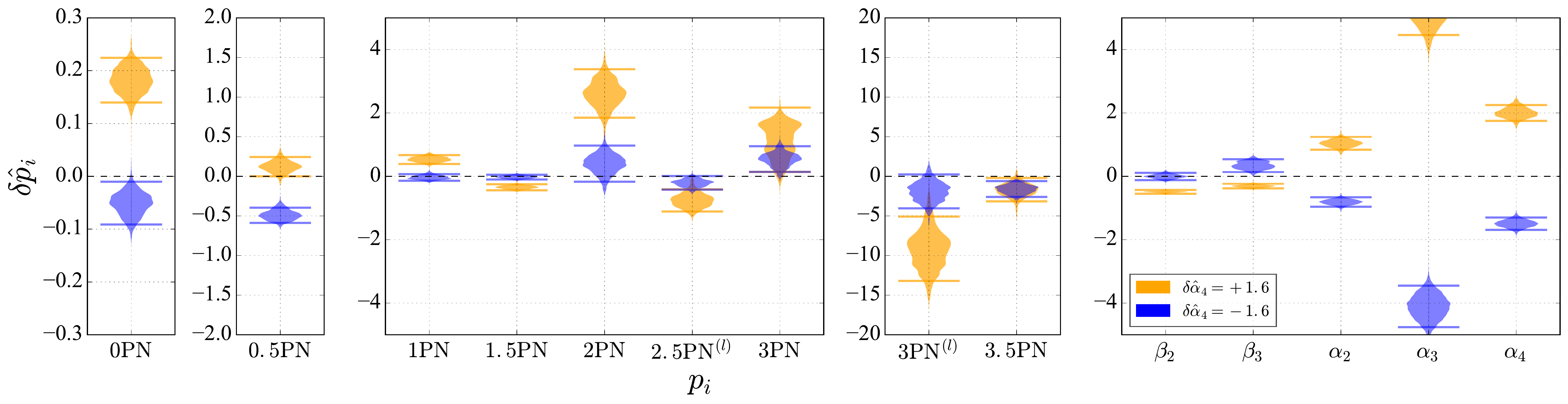}
\caption{Top: Posterior densities for testing parameters for an injection
with $\delta\hat{\alpha}_2 = +1.3$ (orange) and $\delta\hat{\alpha}_2 = -1.3$ (blue).
Bottom: posteriors for an injection with 
$\delta\hat{\alpha}_4 = +1.6$ (orange) and $\delta\hat{\alpha}_4 = -1.6$ (blue). Here too,
in each case the other $\delta\hat{\alpha}_i$ also pick up the GR violation.}
\label{fig:dalphas}
\end{figure*}

The posteriors in Figs.~\ref{fig:dbetas} and \ref{fig:dalphas}, where either an 
intermediate-regime parameter or a merger-ringdown parameter in the signal has a deviation, 
show analogous behavior: For the parameter where the deviation resides, posteriors 
have no support at zero, but this is also the case for at least one other parameter, 
usually one in the same regime. 
  
\subsection{Simulated signals with deviations in multiple coefficients}

Next we consider injections in which all the $\delta\hat{p}_i$ are non-zero 
starting from some PN order. Two scenarios are considered:
\begin{enumerate}
\item All testing parameters starting from 1.5PN have the same fractional 
shifts $\delta\hat{p}_i = 0.5$. This includes the sets 
$\delta\hat{\varphi}_{3,4,5l,6,6l,7}$, $\delta\hat{\beta}_{2,3}$, and
$\delta\hat{\alpha}_{2,3,4}$. 
\item All testing parameters starting from 1.5PN have shifts whose sign 
alternates from one parameter to the next, according to the way they
are correlated: $\delta\hat{\varphi}_{3,5l,6l,7} = -0.4$, and 
$\delta\hat{\varphi}_{4,6} = +0.4$. For the intermediate-regime and
merger-ringdown parameters, we choose
$\delta\hat{\beta}_2 = \delta\hat{\beta}_3 = -0.4$
and $\delta\hat{\alpha}_2 = \delta\hat{\alpha}_3 = \delta\hat{\alpha}_4 = 0.4$. 
\end{enumerate}

\begin{figure*}
\includegraphics[width=\textwidth]{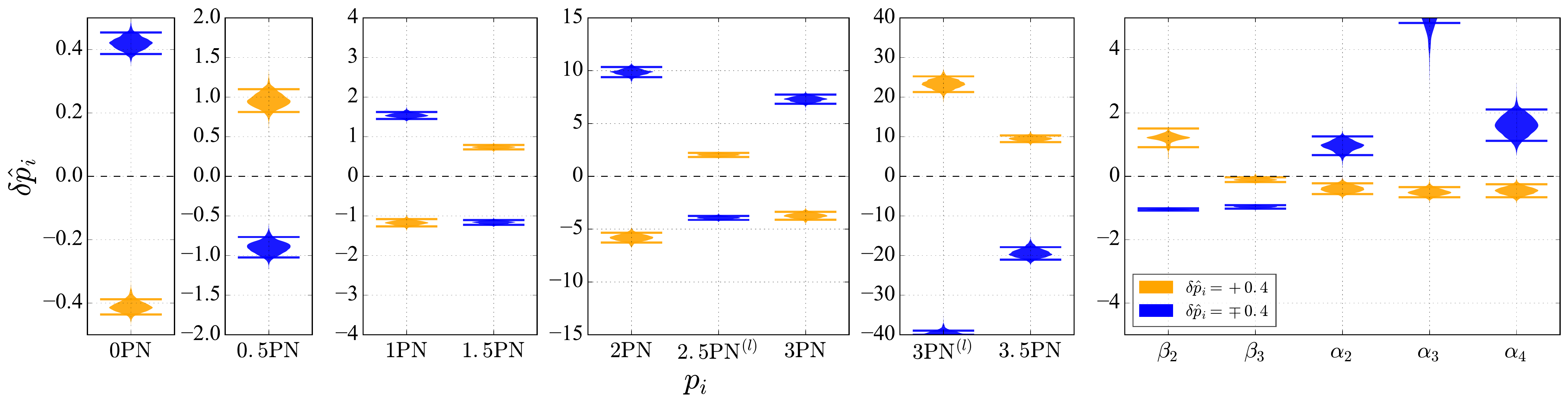}
\caption{Results for injections where all of the signal's testing parameters starting
from 1.5PN have a fractional shift $|\delta\hat{p}_i| = 0.4$,  
in one case all with positive sign (orange), in another case with a sign 
that alternates from one parameter to the next (blue); see the main text for 
details. In both cases the offsets of the posterior densities
follow the way successive PN coefficients are correlated. 
Also note how in both cases all of the $\delta\hat{p}_i$ clearly indicate a GR violation in 
the signal. In fact, from 1.5PN order onwards, the measured violations in PN parameters is \emph{larger} 
than the injected deviation at a given order: Individual testing parameters
will try to accommodate the collective change in the signal resulting 
from the shifts in all the parameters together.}
\label{fig:batches}
\end{figure*}

The results are shown in Fig.~\ref{fig:batches}, and can be summarized as follows:
\begin{itemize}
\item Again strong deviations are picked up even by testing parameters that are
not associated with the violations in the signal; both for the same-sign and 
alternating-sign violations, all of the testing parameters return a posterior density 
function whose support does not contain the GR value of zero.
\item Even in the case where the signs of all the deviations are the same, 
we see alternation in the offsets of the posterior densities for PN parameters, 
following the way they are correlated.
\item  For PN parameters from 1.5PN onwards, 
the measured GR violation is \emph{larger} than the injected deviation; individual 
parameters respond to the collective change in the waveform induced by the shifts in all
of the testing parameters together. 
\end{itemize}
Hence, measuring the $\delta\hat{p}_i$ one by one can enable
the discovery of GR violations also when the signal has multiple $p_i$ that deviate
from their GR values.

\section{Summary and conclusions}

In \cite{TheLIGOScientific:2016src,TheLIGOScientific:2016pea,Abbott:2017vtc}, the detected 
binary black hole signals were analyzed using template waveforms that allow for parameterized
deviations from GR, so as to test the strong-field dynamics of the theory. In this work 
we have introduced reduced-order quadratures that speed up likelihood calculations by factors
of a few to more than a hundred, which will significantly ease the computational burden in 
applying the method to future events. Our chosen waveform model is IMRPhenomPv2, though 
we note that the method used in this paper can in principle also be applied to reduced order 
models for other frequency domain waveforms with parameterized deviations added, such as 
the ones in \cite{Purrer:2014fza,Purrer:2015tud}. We also established the method's robustness through 
$p$-$p$ plots for simulated signals in synthetic Gaussian noise, and by examining the results
for a numerical relativity injection in different stretches of real data from the S6 data set, 
recolored to the Advanced LIGO final design sensitivity. Finally, the sensitivity of the method was
evaluated using both GR injections and injections with GR violations in various parameters.

A range of alternative theories of gravity have been considered, which are often 
characterized by additional charges or coupling constants. The tests presented
here do not easily map to statements about such parameters; putting constraints
on particular alternative theories would require full inspiral-merger-ringdown waveforms
of similar quality as the ones we have for GR. The regular observation of binary black 
hole coalescences will be an incentive for theorists to develop such models.
However, the main aim of the parameterized tests is to perform stringent tests of 
GR itself, and as we have demonstrated, our method provides a reliable and accurate
way of doing this.

Recently a binary neutron star merger was also discovered \cite{TheLIGOScientific:2017qsa}.
Here too the parameterized tests can be applied, although care should be taken so that
the effects of the neutron stars' tidal deformation are not confused with a violation of
GR. This can be done by analyzing the signal up to frequencies of only a few hundred Hertz 
so that tidal effects can be neglected \cite{Li:2011cg,Li:2011vx,Agathos:2013upa}, or by including
tidal deformabilities in the signal. The latter approach has the advantage that the 
entire signal can be used, but there will also be some loss of sensitivity due to the
increased dimensionality of parameter space; which approach will be the most efficient 
is yet to be determined. Especially for these kinds of events, which involve longer 
signals than for binary black holes, it would be beneficial to construct reduced-order
quadratures; this too is left for future work.

\section*{Acknowledgements}

The authors have benefited from discussions with many LIGO Scientific Collaboration (LSC) 
and Virgo Collaboration members. Jeroen Meidam, Ka Wa Tsang, 
Archisman Ghosh, Patricia Schmidt, and Chris Van Den Broeck are supported by the 
research programme of the Netherlands Organisation for Scientific Research (NWO). Michalis 
Agathos acknowledges NWO-Rubicon Grant No.~RG86688. Tjonnie
Li was partially supported by a grant from the Research Grants Council of the Chinese University of 
Hong Kong (Project No. CUHK 24304317), and the Direct Grant for Research from the Research 
Committee of the Chinese University of Hong Kong. John Veitch is supported by UK Science
and Technology Facilities Council (STFC) grant ST/K005014/1. 
Kent Blackburn and Salvatore Vitale acknowledge the support of the National Science 
Foundation and the LIGO Laboratory. LIGO was constructed by the California Institute of 
Technology and Massachusetts Institute of Technology with funding from the 
National Science Foundation and operates under cooperative agreement PHY-0757058.

\bibliographystyle{unsrt}
\bibliography{references}

\end{document}